\documentclass{mn2e}
\input{psfig.sty}
\usepackage{mnras_cite}
\newcommand{\apj}{Astrophys. J.}
\newcommand{\aj}{Astron. J.}
\newcommand{\apjs}{Astrophys. J. Supp.}

\newcommand{\mnras}{Mon. Not. R. Astron. Soc.}

\def\sun{\hbox{$\odot$}}

\newdimen\hssize
\newdimen\hdsize 
\begin{document}

\title[Red and Blue Galaxy Statistics]{A Divided Universe: 
Red and Blue Galaxies and their Preferred Environments}
\author[Cooray]{Asantha Cooray\\ 
Department of Physics and Astronomy, University of California, Irvine, CA 92697\\
E-mail:acooray@uci.edu}

\maketitle

\begin{abstract}
Making use of scaling relations between the central and the total galaxy luminosity 
of a dark matter halo as a function of the halo mass, and the scatter in these relations,
we present an empirical model to describe the luminosity function (LF) of galaxies.
We extend this model to describe relative statistics of  early-type, or red, 
and late-type, or blue, galaxies, with the fraction of early type galaxies at halo centers, relative to the total sample,
 determined only by the halo mass and the same fraction in the case of satellites is taken to be dependent on both the halo mass and
the satellite galaxy luminosity. This simple model describes the conditional luminosity functions,
LF of galaxies as a function of the halo mass, measured with the 2dF galaxy group catalog from
cluster to group mass scales. Given the observational measurements of the LF as a function of the
environment using 2dF, with environment defined by the galaxy overdensity measured over a given volume, we extend our model to
describe environmental luminosity functions. Using 2dF measurements, we extract information related to
 conditional mass function for halos from extreme voids to dense regions in terms of the
galaxy overdensity. We also calculate the probability distribution function of
halo mass, as a function of the galaxy overdensity, and use these probabilities to address preferred 
environments of red and blue galaxies. Our model also allow us to make predictions, for example,
galaxy bias as a function of the galaxy type and luminosity, the void mass function, and
the average galaxy luminosity as a function of the density environment.
The extension of the halo model to construct conditional and environmental luminosity function of galaxies
is a powerful approach in the era of wide-field large scale structure surveys given the ability
to extract information beyond the average luminosity function.
\end{abstract}

\begin{keywords}
large scale structure --- cosmology: observations --- cosmology: theory --- galaxies: clusters:
 general --- galaxies: formation --- galaxies: fundamental parameters ---
intergalactic medium --- galaxies: halos --- methods: statistical

\end{keywords}

\section{Introduction}

An important aspect of understanding underlying astrophysical reasons for
galaxy formation and evolution involves studying 
the relative distribution of red (early-type) and blue (late-type) galaxies,
as a function of the galaxy environment. Observational measurements of
this so-called density-morphology relation suggest evidence that early-type galaxies
are predominantly found in dense environments such as galaxy groups and clusters (Dressler 1980; Goto et al. 2003)
Evidence also suggests that  the formation of early-type galaxies predates the
formation of galaxy clusters (Dressler et al. 1997). In the era of wide-field 
galaxy surveys such as the 2dF Galaxy Redshift Survey (2dFGRS; Colless et al. 2001) or
the Sloan Digital Sky Survey (SDSS; York et al. 2000), 
detailed statistics on galaxy types and their environments allow
one to now construct detailed models related statistics of galaxy types and where these
galaxies are located. These models may then aid in explaining the underlying
reasons for the occurrence of galaxy types and their preferred environments
with initial conditions given by the primordial density fluctuations and
cosmological parameters that determine the expansion. 

Following these lines, numerical and semi-analytical 
models of galaxy formation are generally pursued to model
and understand galaxy statistics including luminosity functions, occurrence of galaxy types,
their spatial distribution, and clustering properties. While initial
conditions and background cosmology are known adequately, these techniques are yet to 
describe statistical measurements of the galaxy distribution with reasonable accuracy. 
The main unknown here comes from our limited understanding of gastrophysics involving how baryons cool to form stars and how the
formation of stars and their evolution, including the stellar end products, may lead to various
feedback processes that affect subsequent starformation. The standard {\it text book} description of
galaxy formation involves  gas first heating to virial temperature during the formation of
dark matter halos and subsequently cooling at halo centers to form stars (Rees \& Ostriker 1977; White \& Rees 1978)
A characteristic scale in galaxy formation is then related to the amount of gas, basically within the
``cooling radius'', that can cool within the Hubble time given a mass scale for the halo.
Semi-analytic models  of galaxy formation (e.g., Benson et al. 2001) and direct
hydrodynamical simulations   of the galaxy distribution (e.g., Kay et al. 2002)  generally over predict the number of
galaxies both at the low-end and the high-end of the galaxy luminosity function.
The bright-end of the LF is always associated with the ``over cooling problem''
in numerical simulations (Balogh et al. 2001), 
where hot gas cools rapidly to form luminous galaxies at halo
centers. The faint-end problem, involving a lack of  faint galaxies in the
observed LF, is generally explained as due to a feed back process
during the era of reionization (Barkana \& Loeb 2001) when primordial galaxies started to form. Models are
generally evoked to expel gas from halos, such as through heating associated with
reionization or a first generation of supernovae (e.g., Bullock et al. 2000; Benson et al. 2002), but the gas expelled 
from  small halos  settle eventually in more massive halos and
cool to form bright central galaxies with luminosities exceeding those observed (e.g., Benson et al. 2003).

Ignoring galaxy growth through continuous cooling of hot gas,
in Cooray \& Milosavljevi\'c (2005a), the central galaxy luminosity growth, as a function of the halo mass
was explained based on a simple description for galaxy merging to the halo center with an efficiency determined
by the dynamical friction alone. These models also lead to a characteristic scale in the galaxy formation
reflected in terms of a flattering of the relation between central galaxy luminosity and halo mass, or $L_c(M)$ relation,
as the halo mass is increased. This characteristic luminosity is associated 
with a critical mass scale  when the dynamical friction time scale becomes close to or exceed the Hubble time.
In Cooray \& Milosavljevi'c (2005b), a model for the LF of galaxies that relied primarily on this relation, 
and to a lesser extent on the
relation between total galaxy luminosity in a given halo and it's halo mass, was used to show that this characteristic
luminosity is same as $L_*$ in the LF, when the LF is   described with the Schechter (1976) form of
$\Phi(L) \propto (L/L_\star)^\alpha \exp(-L/L_\star)$. The success in describing the
LF of galaxies using the $L_c(M)$ relation and it's scatter led to the conclusion that the
$L_*$ in the luminosity function is not a reflection of efficiency associated with gas cooling 
as has been argued in the past 
based on traditional models of galaxy formation dominated by hot gas cooling in dark matter halos (e.g., Dekel 2004). 
The characteristic luminosity is rather due to decreasing efficiency of dissipationless merging of galaxies to a central
galaxy as hierarchical  structure formation builds up massive parent halos. 

Other evidences for a departure from traditional ideas of galaxy formation and evolution comes
from numerical simulations, where some simulations now suggest that gas, as dark matter halos virialize,
never heat to the ``virial'' temperature completely, but rather, shock heating during virialization forms a bimodal
temperature distribution (Dekel \& Birnboim 2004; Keres et al. 2004).
Binney (2004) suggested that only the colder component cools to form a galaxy, while the
hotter component remains at the same temperature. There is a lower characteristic scale in galaxy formation
associated with the mass scale where most gas is never shock heats and remains at the virial temperature, but rapidly
cools at the center to form a galaxy. One-dimensional numerical simulation suggests this lower mass scale is
$(1-6)\times 10^{11}$ M$_{\sun}$ (Dekel 2004). In this description for galaxy formation, galaxies in more massive halos can only 
grow in luminosity only through mergers with other galaxies. The dynamical friction process involved with merging produces
a consistent $L_c(M)$ relation that agrees with observations (Cooray \& Milosavljevi\'c 2005a).
The same relation, when combined with the mass function, leads to the LF, and can be fitted with the Schechter (1976) form
The exponential drop-off of the LF at the  bright-end is reflection of the
scatter in the $L_c(M)$ relation (Cooray \& Milosavljevi\'c 2005b). 

Here, we extend the model of Cooray \& Milosavljevi\'c (2005b) to describe galaxy statistics measured by
the 2dFGRS survey. The main advantage of using 2dFGRS data is the availability of LF measured
as a function of the galaxy type, and the environment, measured in terms of the galaxy overdensity
over the volume determined by size scale of 8 $h^{-1}$ Mpc (Croton et al. 2004). 
The 2dFGRS data also allow measurements of the conditional luminosity function (CLF; Yang et al. 2003b), the 
luminosity function of galaxies as a function of the halo mass (Yang et al, 2005). Our models on the CLF
can be directly compared to these measurements and interesting information on the relative distribution of
galaxy types can be extracted from the data. 

While not complicated as semi-analytical models of galaxy formation,
the empirical modeling approach utilized here has the advantage that one is able to understand 
main ingredients that shape the CLFs  
easily.  The approach builds upon attempts by Yang et al. (2003b) to describe the LF with  CLFs, but
assuming Schechter forms a priori for the CLF, and approaches that are built upon the halo model for
the galaxy distribution (Cooray \& Sheth 2002), but now extended to discuss conditional functions 
(Zheng et al. 2004 with the stellar mass function and Zehavi et al. 2004 in the case of CLFs).
While the halo model has been successfully used to describe statistics of the dark matter field
(such as clustering: Seljak 2000, Peacock \& Smith 2000 or weak lensing: Cooray et al. 2000, Cooray \& Hu 2001),
and basic properties of galaxy clustering (e.g., Scoccimarro et al. 2001),
it is useful to consider more applications of this technique which can provide
information on underlying physics related to the galaxy distribution.

Here, we will separate our discussion to central and satellite galaxies and make few 
assumptions as possible through simple model descriptions. The motivation for the separation of
galaxies to these two divisions are numerous: from the theoretical side,
a better description of the galaxy occupation statistics is obtained when one separates to central and
satellite galaxies (Kravtsov et al. 2004), while from observations, central and satellites galaxies
are known to show different properties, such as color and luminosity (e.g., Berlind et al. 2004).
Our goal here is to consider an analytic description for the LF with
built in model ingredients that recover the observations. We then argue that instead of attempting
to understand mass-averaged statistics such as the LF, it may be best to reproduce main ingredients that
shape CLFs with numerical and semi-analytical models of galaxy formation in order to understand underlying physics.
In the case of CLFs, we will argue that the main
ingredient is the $L_c(M)$ relation, and in the case of galaxy types, a model on the fraction
of early-type and late-type galaxies as a function of the halo mass. If these relations and their observed scatter can be
explained with simple physics, then it is guaranteed that the LF would be recovered.

To model the environmental LF (Croton et al. 2004), we need information on  the mass function of
 dark matter halos that corresponds to the environment of interest, whether it is a void or a dense region.
Since this information is not directly available under a simple model (see, for example,
Mo et al. 2004 and the approach there that utilized numerical simulations), here
we make use of the observed measurements to extract information on these conditional mass functions.
These mass functions, as well as related probabilities, then provide us with general
information on how blue and red galaxies are distributed in the Universe.

The paper is organized as follows: In the next
 section, we will outline the basic ingredients in the empirical model  for CLFs
and present a comparison to measurements using 2dF data (from Yang et al. 2005), 
both in terms of average number of galaxies and in terms of galaxy types.
In Section~3, we will describe the LF and in Section~4 the environmental LF based on the conditional mass function.
Using data from Croton et al. (2004), we will extract information on the
conditional mass function and various statistical measurements related to the
relative distribution of early- and late-type galaxies. We will also study the void mass function and
compare with predictions in the literature. We conclude with a brief discussion of our main results in
\S~5. Throughout the paper, we assume the concordance cosmological parameters consistent with WMAP data
(Spergel et al. 2003). Throughout this paper, to be consistent with observations,
we take the scaled Hubble constant to be $h=1$, in units of 100 km s$^{-1}$ Mpc$^{-1}$.

\begin{figure}
\centerline{\psfig{file=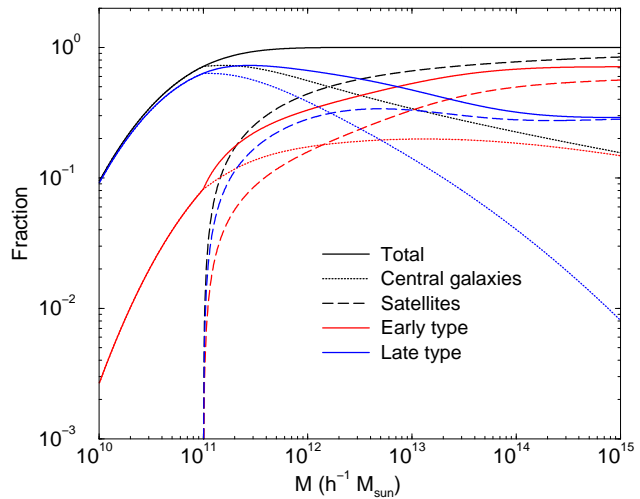,width=\hssize,angle=-90}}
\caption{Various galaxy fractions (central, satellites, late-types and early-types)
as a function of the halo mass. The central galaxy fraction falls below 1 at the
low mass end as we account for the fact that galaxy formation is inefficient at this low end
with the function $f_{\rm c}(M)$ such that not all dark matter halos host a galaxy at the 
low-end of the mass function. The early- and late-type fraction of satellite galaxies also dependent on the
galaxy luminosity. Here, we show the fractions for satellites with luminosity of $10^{10}$ L$_{\sun}$.}
\end{figure}

\section{Conditional Luminosity Function: Empirical Model}

In order to construct the luminosity function (LF), we follow Cooray \& Milosavljevi\'c (2005b).
The conditional luminosity function (CLF), denoted by $\Phi(L|M)$, 
is the average number of galaxies with luminosities between $L$ and $L+dL$ that 
reside in halos of mass $M$ (Yang et al. 2003b).  First, we separate the CLF into terms associated with central and 
satellite galaxies,  such that
\begin{eqnarray}
\Phi(L|M)&=&\Phi_{\rm c}(L|M)+\Phi_{\rm s}(L|M) \nonumber \\
\Phi_{\rm c}(L|M)  &=& \frac{f_{\rm c}(M)}{\sqrt{2 \pi} \ln(10) \Sigma L} \exp \left\{-\frac{\log_{10} [L /L_{\rm c}(M)]^2}{2 \Sigma^2}\right\}  \nonumber \\
\Phi_{\rm s}(L|M) &=& A(M) L^{\gamma(M)} f_{\rm s}(L)\, .
\label{eqn:clf}
\end{eqnarray}

Here $L_c(M)$ is the relation between central galaxy luminosity of a given dark matter halo and it's halo mass, while $\ln(10) \Sigma$ is the
dispersion in this relation. The central galaxy CLF takes a log-normal form, while the satellite galaxy CLF takes a power-law form in luminosity.
Such a separation describes the LF best, with an overall better fit to the data in the K-band as explored by
Cooray \& Milosavljevi\'c (2005b). While a previous attempt to describe CLFs, as appropriate for 2dFGRS, involved a priori assumed
Schechter (1976) forms, we believe the description here is more appropriate. Our motivation for log-normal distributions also come
from measured conditional LFs, such as galaxy cluster LFs including bright galaxies, where data do require an additional
log-normal component in addition to the Schechter (1976) form (Trentham \& Tully 2002). Similarly, the stellar mass
function, as a function of halos mass in semi-analytical models, is best described with a log-normal component for the central galaxies
(Zheng et al. 2004).

In the next few subsections, we will describe in detail other parameters associated with the CLF and how numerical values for these
parameters are obtained. First, we discuss the CLF of central galaxies and then move on to discuss satellites. We will end this section with
a comparison to CLFs measured in 2dFGRS by Yang et al. (2005).

\begin{figure}
\centerline{\psfig{file=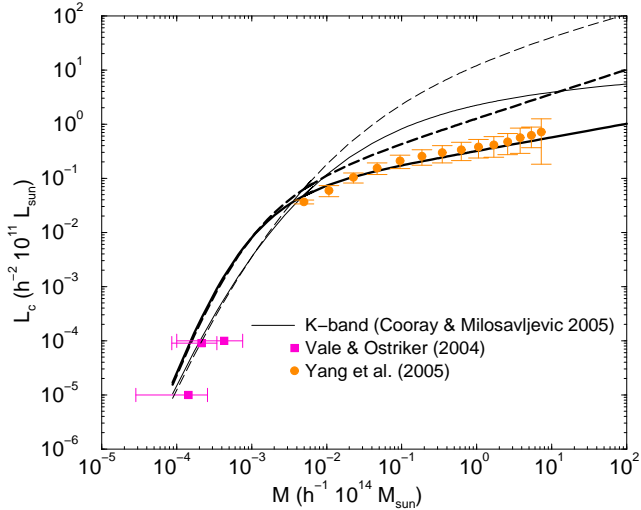,width=\hssize,angle=-90}}
\caption{Central galaxy luminosity as a function of the
halo mass as appropriate for 2dFGRS $b_J$-band. 
The data points at the low mass end are from Vale \& Ostriker (2004; {\it filled diamonds})
while at the high end are from Yang et al. (2005). The solid curve is
the relation obtained in Vale \& Ostriker (2004) by converting the 2dFGRS luminosity function to
extract the plotted relation based on the sub-halo mass function. We use this relation to construct our
the CLF, and include a scatter in this relation in our description of central galaxy CLFs.
The dashed-line is the total luminosity in a given halo. This relation is needed to determine the
CLF of satellites. For comparison,  we also show the central and total-luminosity as a function of halo mass
in the K-band, as obtained by Cooray \& Milosavljevi\'c (2005b) based on a variety of observational measurements
on the central and total galaxy luminosity as a function of halo mass (e.g., with weak lensing:
Yang et al. 2003a; direct measurements: Lin \& Mohr 2004, Lin et al. 2004). 
The Yang et al. (2005) data are direct measurements of
central luminosities based on the 2dFGRS galaxy group catalog. The remarkable agreement between these measurements and the
relation extracted by Vale \& Ostriker (2004) suggests that the $L_c(M)$ relation, as appropriate for the $b_J$-band 
is well established at halo mass scales above $\sim 10^{12}$ M$_{\sun}$.
}
\end{figure}

\subsection{Central Galaxies}

In our description for CLFs (Eq.~1), central galaxies have a log-normal distribution in 
luminosity with a mean determined by the $L_{\rm c}(M)$ relation.  The scatter in the
$L_c(M)$ relation is captured through the dispersion $\Sigma$ of the log-normal distribution.
In Cooray \& Milosavljevi\'c (2005b), we found $\Sigma \sim 0.23$ to describe the field-galaxy luminosity function in the K-band
(Huang et al. 2003). Here, to describe 2dFGRS data in the $b_J$-band, we found a lower
dispersion, with $\Sigma \sim 0.17$. 
While the exact reason for differences between the dispersion at two wavelengths
is not understood, since $\Sigma$ reflects the scatter in the $L_c(M)$ relation, we expect
this relation for luminosities measured in the $b_J$ band to have less scatter than in the K-band.
A comparison of Yang et al. (2005) data shown in Fig.~2 and Lin et al. (2004) data shown in Fig.~1(a) of
Cooray \& Milosavljevi\'c (2005b) suggests this may be the case, but reasons for the difference in
scatter is yet to be understood. Incidently, a value for the dispersion of 0.17 is
in good agreement with the value of 0.168 found for the dispersion of
central galaxy luminosities by Yang et al. (2003b), where these authors used a completely different parameterization for the
CLF then the one described here.

In Eq.~\ref{eqn:clf}, the normalization factor $f_{\rm c}(M)$ in the
central galaxy CLF captures the efficiency for galaxy formation as a function of the halo.
In Cooray \& Milosavljevi\'c (2005), this was set by requiring $\int\Phi(L|M)LdL$ equals the average
 total luminosity of galaxies $L_{\rm tot}(M)$ in a halo of mass $M$.
This condition does not include the fact that
at low mass halos, galaxy formation is inefficient and not all dark matter  halos host a galaxy.
This is equivalent to modifying the halo mass function at the low-mass end to select only halos that host a galaxy.
Motivated by the halo occupation number models for central galaxies (e.g., Kravtsov et al. 2004),
where not all low mass halos occupy galaxies, 
to fit the low-end luminosity data of the 2dFGRS galaxy LF, we allow a description of the form
\begin{equation}
f_{\rm c}(M) = \frac{1}{2}\left[1+{\rm erf}\left(\frac{\log(M)-\log(M_{\rm min})}{\sigma}\right)\right] \, ,
\label{eqn:fcm}
\end{equation}
with parameters $M_{\rm min}=5 \times 10^{10}$ M$_{\rm sun}$, and $\sigma=0.75$.
These parameters were determined by comparing the 2dFGRS LF at the low-end as discussed in Section~3.
This efficiency function is such that it is 0.1 when $M \sim 10^{10}$ M$_{\sun}$, but
is unity when $M >$ few times 10$^{11}$ M$_{\sun}$. When describing the environmental
LFs, we will continue to use this form. 

\begin{figure*}
\centerline{\psfig{file=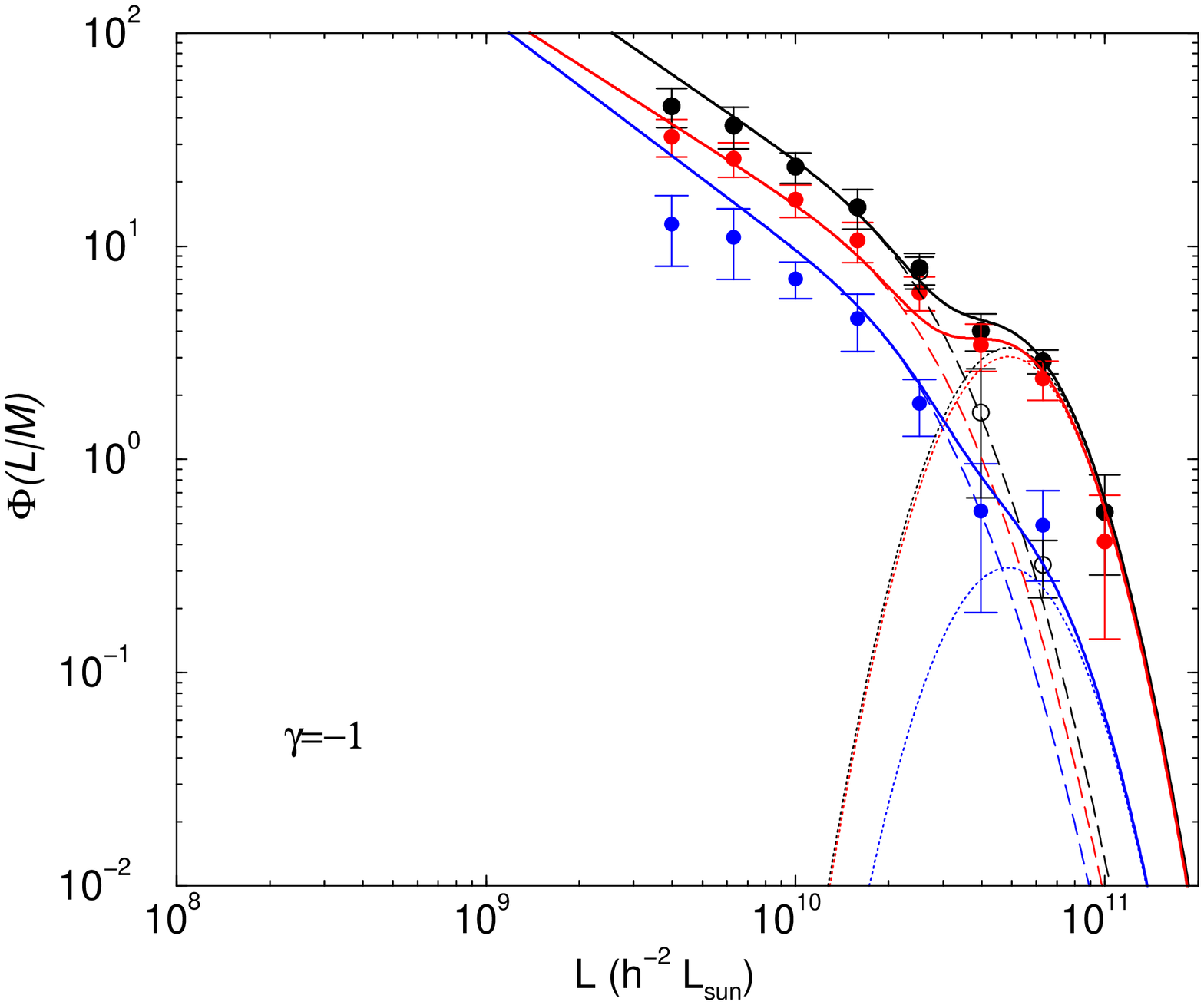,width=\hssize,angle=-90}
\psfig{file=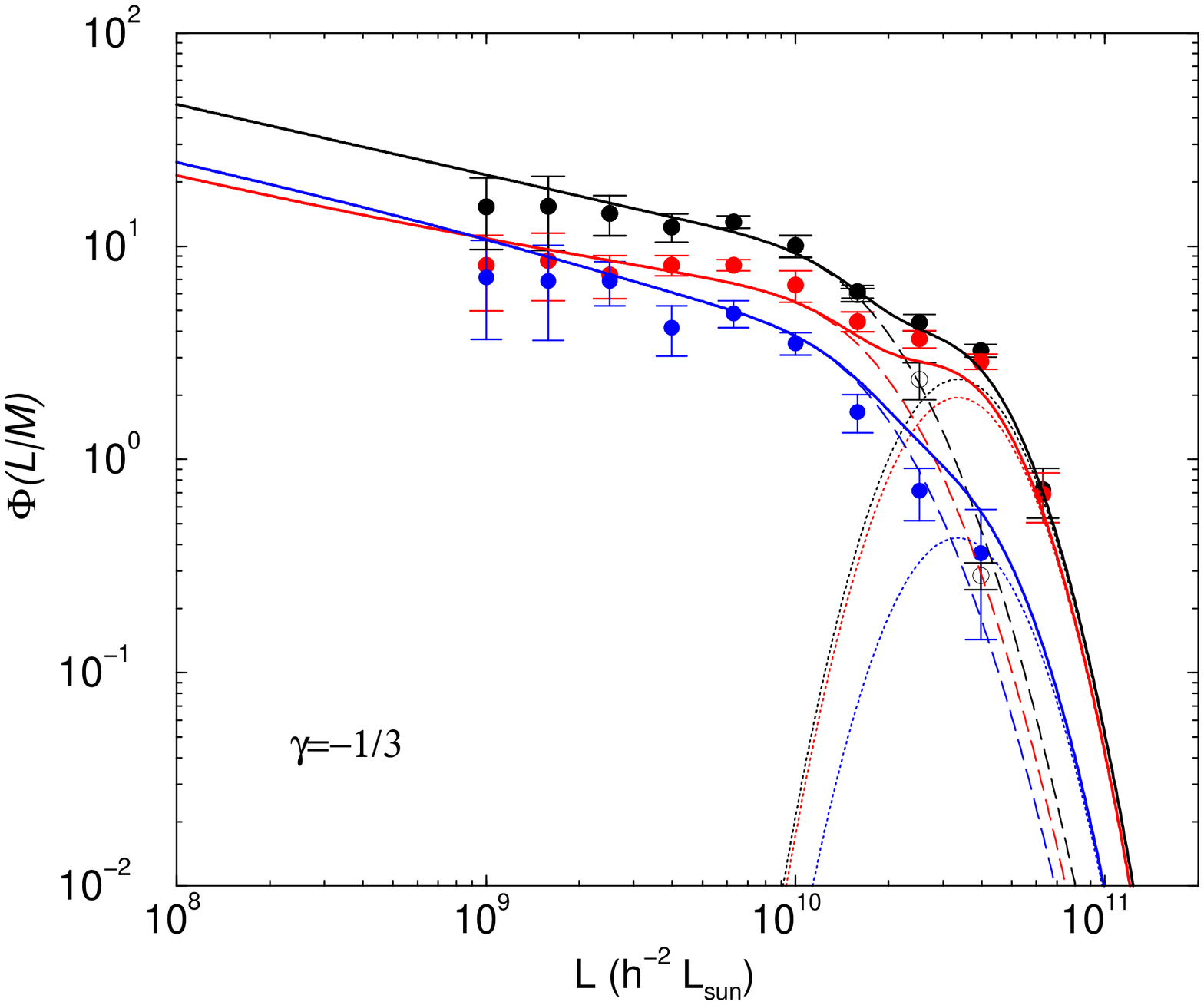,width=\hssize,angle=-90}}
\centerline{\psfig{file=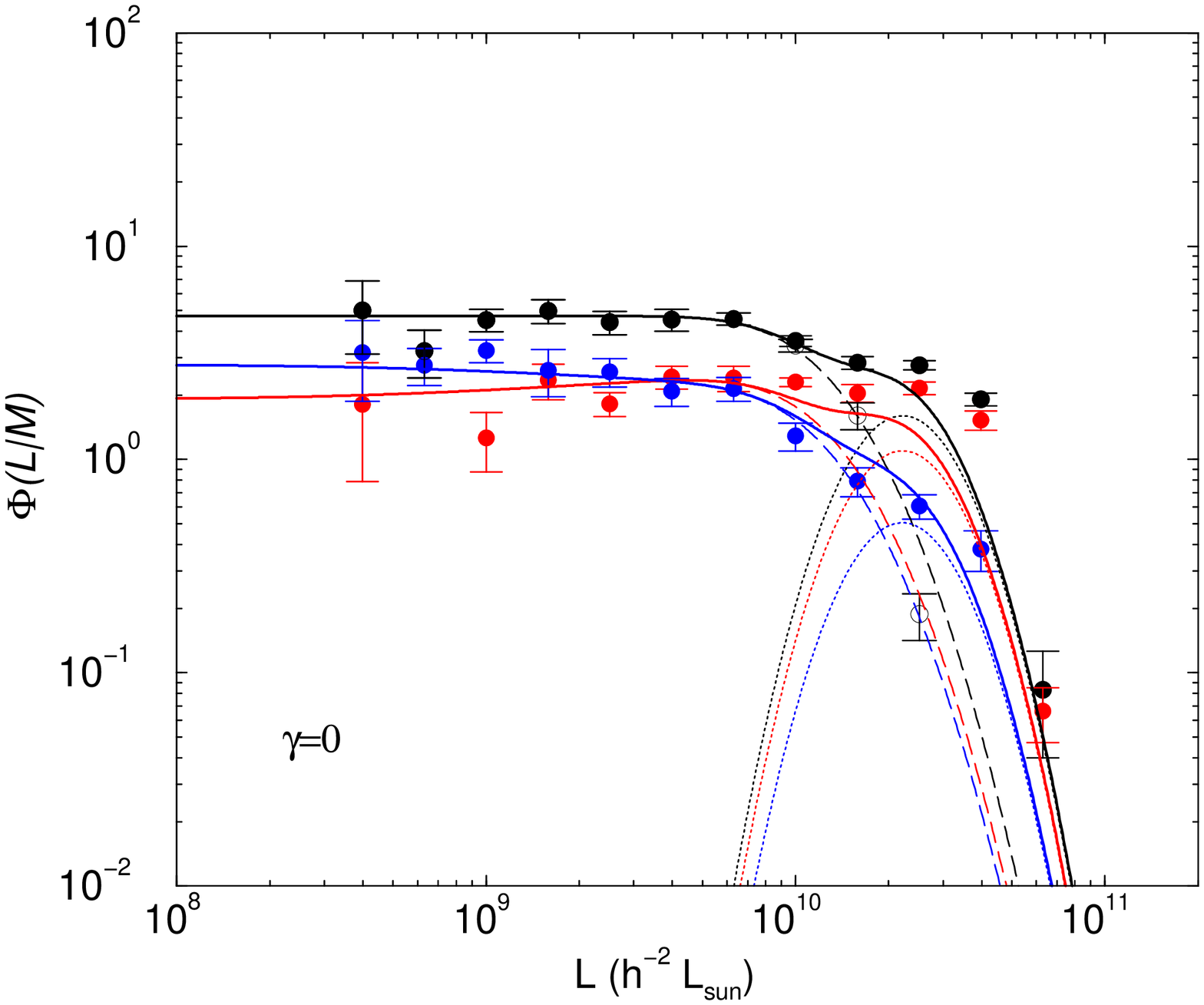,width=\hssize,angle=-90}
\psfig{file=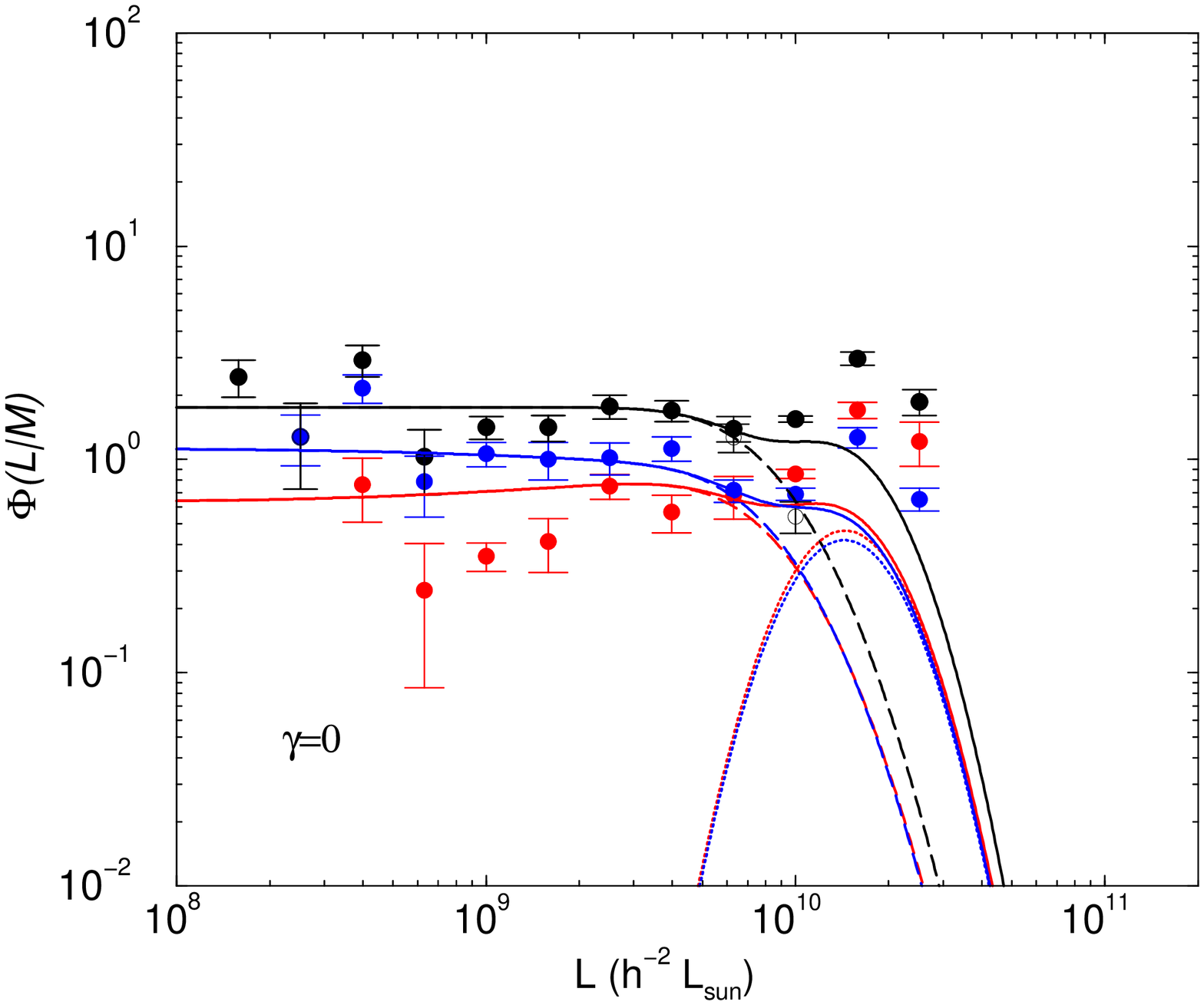,width=\hssize,angle=-90}}
\caption{The conditional luminosity functions $\Phi(L|M)$ for a variety of mass bins as labeled in each of the panels
(with ranges written in logarithmic values of the halo mass).
The data are measurements by Yang et al. (2005) using the galaxy group catalog of the 2dFGRS survey.
To describe conditional luminosity functions, $\gamma(M)$ varies from $-1$ at the high mass end to 0 at the low mass end.
The comparison suggests the evidence of log-normal description for central galaxies and provides
an overall better fit to the data than the description considered by Yang et al. (2005) based
on a priori assumed Schechter (1976) functions for the CLF. Our model, however, under predicts the luminosity of
central galaxies in halos of mass $\sim$ 10$^{12}$ M$_{\sun}$.   While parameters
in our model can be modified to fit data better, especially a modification to the Vale \& Ostriker (2004)
$L_c(M)$ relation, we have not pursued this possibility as the
overall description is adequate enough for the statistical study considered here. Furthermore, it is
also unclear to what extent halo mass estimates at the low-mass end of the 2dFGRS galaxy group catalog can be trusted
and the extent to which selection biases contaminate the CLF determinations at this low mass end.}
\end{figure*}

\subsection{Satellites}

For satellites, the normalization $A(M)$ of the satellite CLF can be obtained by defining $L_{\rm s}(M)\equiv L_{\rm tot}(M)-L_{\rm c}(M)$ and requiring that $L_{\rm s}(M)=\int_{L_{\rm min}}^{L_{\rm max}} \Phi_{\rm s}(L|M)LdL$ with $f_{\rm s}(L)=1$, where 
the minimum luminosity  of a satellite is $L_{\rm min}$.  
In the luminosity ranges of interest, our CLFs are mostly independent of the exact value assumed for $L_{\rm min}$, as long as it lies in the range $(10^6-10^8)L_{\sun}$.  For the maximum luminosity of satellites, following the
result found in Cooray \& Milosavljevi\'c (2005b), by comparing predictions to the
K-band cluster LF of Lin \& Mohr (2004),  we set $L_{\rm max}=L_{\rm c}/2$.
A comparison to 2dFGRS CLFs  as measured by Yang et al. (2005), however,
suggested that such a sharp cut-off is inconsistent
and that to account for scatter in the total galaxy
luminosity, as a function of the halo mass, one must allow for a distribution in
$L_{\rm max}$. Instead of additional numerical integrals, we allow for a luminosity dependence with the introduction of
$f_{\rm s}(L)$ centered around the maximum luminosity of satellites such that
$\Phi_{\rm s}(L|M)$ does not go to zero rapidly at $L_{\rm max}$. By a comparison to the data,
we again found a log-normal description with
\begin{equation}
f_{\rm s}(L) = \frac{1}{2}\left[1+{\rm erf}\left(\frac{\log(L_{\rm c}/2.0)-\log(L)}{\sigma_{s}}\right)\right] \, ,
\end{equation}
where $\sigma_{s}=0.3$. The description here is such that $f_{\rm L} =1$ when $L < L_{\rm mac}=L_{\rm c}/2$,
but falls to zero at a luminosity beyond $L_{\rm c}/2$ avoiding the 
sharp drop-off at $L_{\rm c}/2$ with $f_{\rm s}(L)=1$.  When model fitting
the LF, or the LF as a function of the environment, this description is unimportant
as the central galaxies dominate galaxy statistics. This is due to the fact that,
as discussed in Cooray \& Milosavljevi\'c (2005b), the LF is dominated by central galaxies
instead of satellites, which has also been noted by Zheng et al. (2004) when describing the
stellar baryonic mass function. Though $f_{\rm s}(L)$ does not
matter for the LF, it is important when comparing the CLF of galaxies in halos with a narrow
mass range and when the CLF is measured with central galaxies removed from the data.
With regards to satellites, note that $L_{\rm tot}(M)$ equals $L_{\rm c}(M)$ 
for $M<10^{11}M_{\sun}$, and thus $L_{\rm s}(M)=0$ at low halo masses. Such halos only have a single galaxy,
at the center of the halo.

In Eq.~\ref{eqn:clf}, $\gamma(M)$ is taken to be a function of the halo mass.
In Cooray \& Milosavljevi\'c (2005b), we found $\gamma \sim -1$ based on model fits to the
cluster LF of Lin et al. (2004), and we used that value in all mass scales there when
modeling the K-band LF. Here, based on a comparison to CLFs measured in the 2dFGRS galaxy group catalog, 
we find that $\gamma(M)$ is, in fact, a function of mass 
that varies from -1 at cluster scales to 0 at masses corresponding to groups with 
few galaxies.  A similar mass dependent variations on the faint-end slope, parameterized by 
$\alpha(M)$ in Schechter function forms for the CLF was found by van den Bosch et al. (2005)
when model fitting the CLF using 2dF LF and luminosity-dependent galaxy bias.
While the differences associated with variations to the slope, as a function of mass,
are significant when CLFs between galaxy clusters and poor groups are considered, in the
case of the LF, which averages over CLFs of various mass scales,
the difference resulting from either assuming a mass-dependent slope for $\gamma$ or
an average value, such as $\gamma=-0.5$, is minor. Again, this is due to the fact that
the LF is dominated by central galaxies rather than the satellites.
Thus, we will set $\gamma(M)\sim -0.5$ when calculating the LF and the environmental LFs
in the present paper, but when describing the CLFs of Yang et al. (2005), we allow for a mass dependence for $\gamma$
(see, Figure~3).

\begin{figure*}
\centerline{
\psfig{file=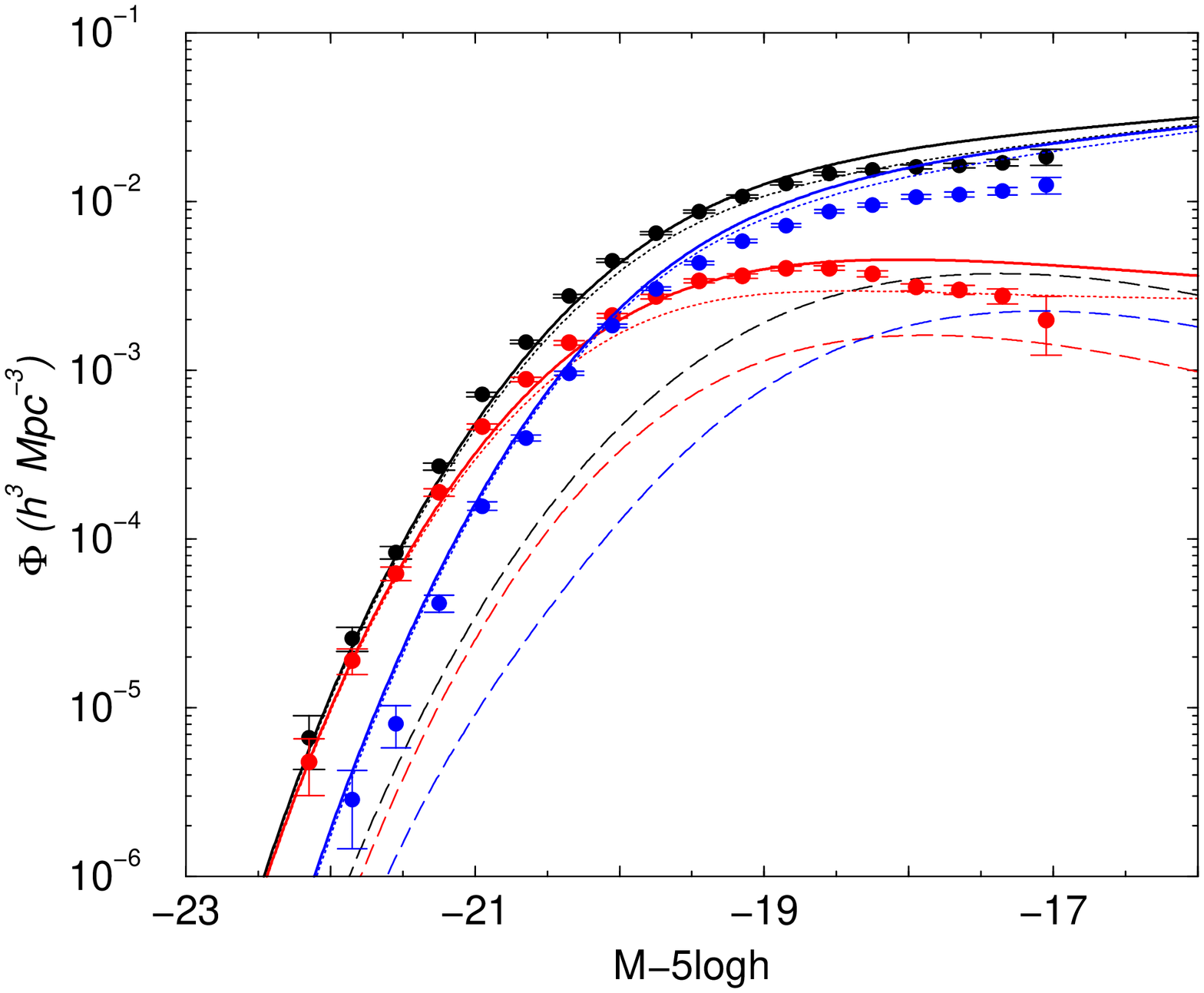,width=\hssize,angle=-90}
\psfig{file=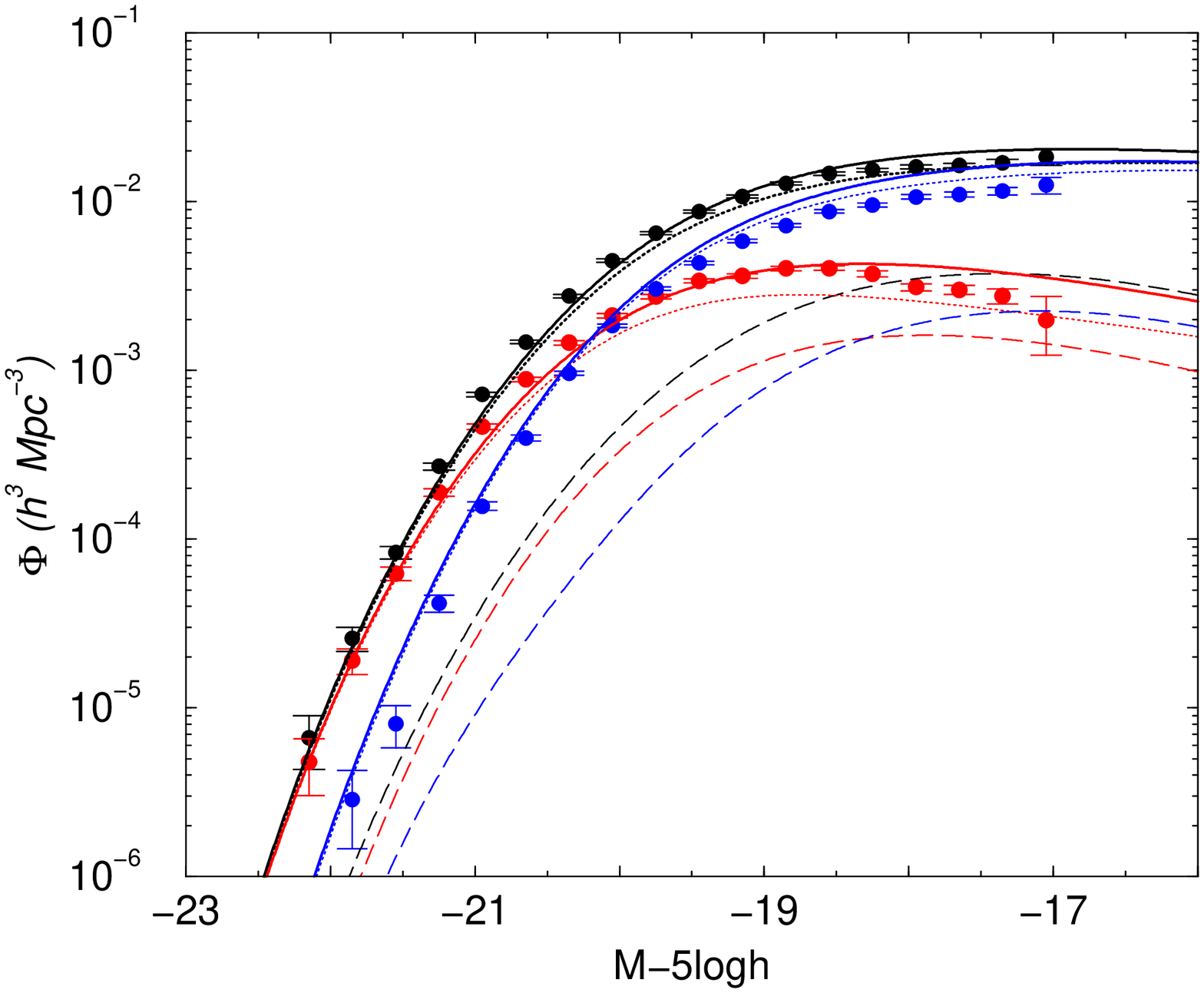,width=\hssize,angle=-90}}
\caption{ The galaxy LF in the 2dFGRS survey. The measured data points are from Croton et al. (2004).
In addition to the total luminosity function, we also show the
early (red lines) and late-type (blue lines) LFs. The dotted lines are the central galaxies while the dashed lines
are for satellites. In (a), we set $f_{\rm c}(M)=1$. This results in a faint-end slope
of $-\sim 1.3$, which is steeper than the measured value of $\sim -1.05$. In order
to account for the flattening, in (b), we allow for $f_{\rm c}(M)$ to be mass dependent
with the form given in equation~\ref{eqn:fcm}.}
\end{figure*}

\subsection{Galaxy types}

Here, in addition to the 2dFGRS LF, we will also model the LF of galaxy types, broadly  divided in to two classes
involving red, or early-type, and blue, or late-type, galaxies. 
A previous modeling of galaxy types using CLFs is described in van den Bosch et al. (2003); our approach
differs because of the overall division of the sample to central and satellite galaxies.
Thus, the division to two types applies to these two components,
separately, in a given halo. As it is clear, the CLFs we have developed facilitate this easy separation.

As we find later by comparing to 2dFGRS data,
central galaxies tend to be early-type when found in massive halos, corresponding to groups and clusters, but late-type
when in low mass halos with a few or no satellites. Thus, the division of central galaxies to the two types
can simply be described as a function of mass. While we have extracted this division in a function form here, we 
have not investigated the underlying reasons how galaxy types in the center of halos change from primarily late to
early type, as the halo mass function increases. The form is given and it remains to be seen if this
analytical description is reproduced in numerical or semi-analytic models of galaxy formation or not.

In the case of satellites, a comparison to the 2dFGRS CLFs measured by Yang et al. (2005)
shows that the division to early- and late-type galaxies is simply not a function of mass alone, 
but rather a function of both mass and luminosity of the galaxy. For example, in low mass halos corresponding to
galaxy groups, the low luminous satellites with luminosities less than 10$^9$ L$_{\sun}$ 
tend to be mostly late type  or blue galaxies, while the bright satellites, with
luminosities around 10$^{10}$ L$_{\sun}$ are dominated by  early type or red galaxies. 

To describe these general behaviors, we consider the division of
central and satellite conditional luminosity functions to early- and late-type separately, and 
write
\begin{eqnarray}
\Phi_{\rm early-cen}(L|M) &=& \Phi_{\rm c}(L|M) f_{\rm early-cen}(M) \nonumber \\
\Phi_{\rm early-sat}(L|M) &=& \Phi_{\rm s}(L|M) f_{\rm early-sat}(M,L) \, ,
\end{eqnarray}
where the two functions that divide between early- and late-types are taken to be
functions of mass, in the case of central galaxies,
and both mass and luminosity in the case of satellites.  Note that these fractions are defined
with respect to the total galaxy number of a halo. Since the early- and  late-type
fractions should sum to unity, late-type fractions are simply $[1-f_{\rm early-cen}(M)]$
and $[1-f_{\rm early-sat}(M,L)]$ for central and satellite galaxies, respectively. Given this simple dependence,
we do not write late-type fractions separately.

In the case of central galaxies, we assume that there is
a smooth transition between a dominant fraction of late-type to a dominant early-type fraction, as a function of the halo mass,
as the halo mass is increased. A description that fits the 2dFGRS CLFs of Yang et al. (2005) was
\begin{equation}
f_{\rm early-cen}(M) = \frac{1}{2}\left[1+{\rm erf}\left(\frac{\log(M)-\log(M_{\rm cen})}{\sigma_{\rm early-cen}}\right)\right] \, ,
\end{equation}
with $M_{\rm cen} = 5 \times 10^{12}$ M$_{\sun}$ and $\sigma_{\rm early-cen}=2.0$;
As the halo mass increased, a large width for the transition from predominantly late-type to early-type
galaxies in the centers of massive halos results in an early-type fraction that never falls to unity or zero
either of the two ends in the mass ranges of interest.

For satellites, at high mass halos, we found early type fraction to be 
roughly two-thirds  while at the low mass end, this fraction decreases
to around one-third. At the low end of halo masses where satellites are found, however, this fraction is 
luminosity-dependent. The model that describes this behavior is
\begin{eqnarray}
f_{\rm early-sat}(M,L) =  \frac{1}{6}g(M) +\frac{1}{6}h(L)+\frac{1}{3} \, ,
\end{eqnarray}
where,
\begin{eqnarray}
g(M) &=& \frac{1}{2}\left[1+{\rm erf}\left(\frac{\log(M)-\log(M_{\rm sat})}{\sigma_{\rm sat}}\right)\right] \nonumber \\
h(L) &=& \frac{1}{2}\left[1+{\rm erf}\left(\frac{\log(L)-\log(L_{\rm sat})}{\sigma_{\rm sat}}\right)\right] \, ,
\end{eqnarray}
with $M_{\rm sat}=3 \times 10^{13}$ M$_{\sun}$, $L_{sat} = 3 \times 10^{9}$ $L_{\rm sun}$, and $\sigma_{\rm sat} = 1$. 
This function varies between $1/3$ and $2/3$ when low-mass, low-luminosities to high-mass, high-luminosities.
Note that satellites are subhalos that have merged with a central halo. Thus, before these galaxies became satellites. they were
in fact central galaxies. Thus, the fact that the fraction of early-to-late satellite galaxies is both mass and luminosity
dependent should not be considered a drawback in this description or the halo approach to galaxy statistics in general.
In fact, a model where redshift dependences are also included, including a merging hierarchy, one may able to
start with simple a description for early-to-late type galaxy fraction that is mass dependent alone and
understand how the merging of these galaxies result in the fractional dependence of early galaxies, say, as satellites
in  massive dark matter halos. Clearly, such work follows underlying motivations of semi-analytical models of galaxy
formation. Here, we provide the relations that needed to be explored in such an approach.
 
In Figure~1, as a summary, we show, as a function of the halo mass various fractions encountered in our model.
Note that the central galaxy fraction falls below unity at halo masses below 10$^{11}$ M$_{\sun}$, which is due to
the efficiency factor we included to account for the fact that not all halos at the low-end may host a galaxy.
Setting $f_c(M)=1$ results in an over prediction for the abundance of galaxies at that low-luminosity end.
We discuss this in the context of modeling the 2dFGRS galaxy LF (Section~3).

\subsection{Central and total-luminosity relation}

The main ingredient in the modeling the LF using this empirical approach is the $L_{\rm c}(M)$ relation.
As discussed in Cooray \& Milosavljevi\'c (2005b), the shape of the $L_{\rm c}(M)$ relation determine the shape of
the LF; The slope of this relation is directly reflected in the faint-end slope of the LF, while the scatter of this
relation determines the exponential-like drop off of the LF at the bright-end.

For $L_{\rm c}(M)$ relation, here we make use of the suggested relation in Vale \& Ostriker (2004).
These authors established this relation by inverting the 2dFGRS luminosity function given a analytical description for the sub-halo
mass function of the Universe (e.g., De Lucia et al. 2004; Oguri \& Lee 2004). 
The relation is described with a general fitting formula given by
\begin{equation}
\label{eq:fitting}
L(M) = L_0 \frac{(M/M_1)^{a}}{[b+(M/M_1)^{cd}]^{1/d}}\, .
\end{equation}
For central galaxy luminosities, the parameters are $L_0=5.7\times10^{9} L_{\sun}$, $M_1=10^{11} M_{\sun}$,
$a=4.0$, $b=0.57$, $c=3.72$, and $d=0.23$ (Vale \& Ostriker 2004). 
These values are different from Cooray \& Milosavljevi\'c (2005b) since we use 
K-band luminosities there, while the relation given in this paper is expected to describe the 2dFGRS data adequately, 
as it is extracted from 2dFGRS LF in the $b_J$-band of Norberg et al. (2002a). 

For the total galaxy luminosity, as a function of the halo mass, 
we also use the fitting formula in equation (\ref{eq:fitting}), but with 
$c=3.57$, which we picked based on model fits to the 2dFGRS CLFs of Yang et al. (2005).
At the massive end, the total luminosity can alternatively be 
described by a power-law or a double power-law with the break around $\sim 3.5 \times 10^{13}$ M$_{\sun}$ (Vale \& Ostriker 2004). 
The constructed LF does not change if power-law behavior is 
enforced at the high-end of the halo masses. 
This is because the average LF is dominated by central galaxies on any scale. 
The overall shape of the LF is {\it strongly} sensitive to the shape of the $L_{\rm c}$--$M$ relation,
and it's scatter, and less on details related to the $L_{\rm tot}--M$ relation.

In Figure~2, we show the two relations used for central galaxy and total galaxy luminosity of a given halo, on
average, as a function of the halo mass. Incidently, we found the relation extracted by
Vale \& Ostriker (2004) to be in good agreement with  direct measurements of central galaxy luminosities in the 2dF galaxy
group catalog by Yang et al. (2005). Similarly, the $L_{\rm tot}--M$ relation based on
Vale \& Ostriker (2004) agrees with total luminosity measurements independently obtained by a different
technique, involving clustering relation, in van den Bosch et al. (2005). 
In fact, these agreement are remarkable and suggest that we have a good starting point to
build up a model for the galaxy LF both as a function of the environment and galaxy type.

\subsection{Conditional Luminosity Functions: A Comparison to 2dF}

Now that we have an analytical description for the CLF with parameters determined either by results already in the
literature, such as the $L_{\rm c}(M)$ from Vale \& Ostriker (2004), or based on model fitting the data, we can
discuss how well our models fits the 2dF CLFs of Yang et al. (2005). These CLFs were previously modeled with a priori
assumed Schechter (1976) functions following the models
in Yang et al. (2003b), but here we make use of the log-normal description for central galaxies
and power-laws for satellite galaxies to build the CLF.

In Figure~3, we show a comparison of our CLFs to those extracted from 2dF galaxy group catalog by Yang et al. (2005).
We show our models in mass ranges they measured the CLFs. These CLF model fits can in fact be compared with
Figures~9 and 11 in Yang et al. (2005). The models fits generally support our log-normal description for central galaxies
and the separate description of power-law for satellite galaxies.
The model fits require that the slope of the power-law be  $\gamma \sim -1$ in cluster scales,
as found by Cooray \& Milosavljevi\'c (2005b) by comparing to the cluster luminosity function of Lin et al. (2004)
in the K-band, but flattens to $\gamma \sim 0$ at galaxy group scales. The transition from late-to-early type is
also adequately modeled with our simple description, except that we find our models to under predict the
number of bright, and potentially central, galaxies at poor galaxy group mass scales. This 
difference may be modeled by updating the $L_c(M)$ at these mass scales, but given the overall adequate description,
and the fact that we wanted to build this model with the least number of parameter variations as possible but 
by using existing results from the literature (such as from Vale \& Ostriker 2004), we have not pursued
such a possibility here. It is also not clear how well masses have been estimated by Yang et al. (2005)
for low mass galaxy groups where few galaxies are found in each group. 

As discussed in Yang et al. (2005) for 2dF b-band data, and
in Cooray \& Milosavljevi\'c (2005b) for the K-band data, 
the CLF represents galaxy statistics better than the LF when wide-field data sets are 
available in which redshifts are measured for tens of thousands of galaxies.
While Yang et al. (2003b) described CLFs using a priori assumption that
$\Phi(L|M)$ is given by the Schechter (1976) function 
with no separation to central and satellite galaxies, as it is clear from Figure~3,
our description involving log-normal distribution for central galaxies and a power-law
for satellite galaxies may provide a better description. However, the peak of
galaxies at the bright-end may be due to a problem associated with mass assignment in the
2dFGRS galaxy group catalog by Yang et al. (2005; van den Bosch, private communication).
The same reasons could also explain the apparent increase in bright galaxies at low mass halos, such
as poor groups, when compared to our model predictions. 
Even if improvements are minor, when compared CLFs based
on the Schechter form used in  Yang et al. (2005), 
the method suggested here based on a division of the galaxy sample to central and satellite galaxies is more physical.
As described, the model considered also provides a more useful approach to divide
galaxies to galaxy types, which is advantageous since we are trying to get
an understanding of how galaxy types are distributed in varying dark matter halo masses.

\begin{figure*}
\centerline{
 \psfig{file=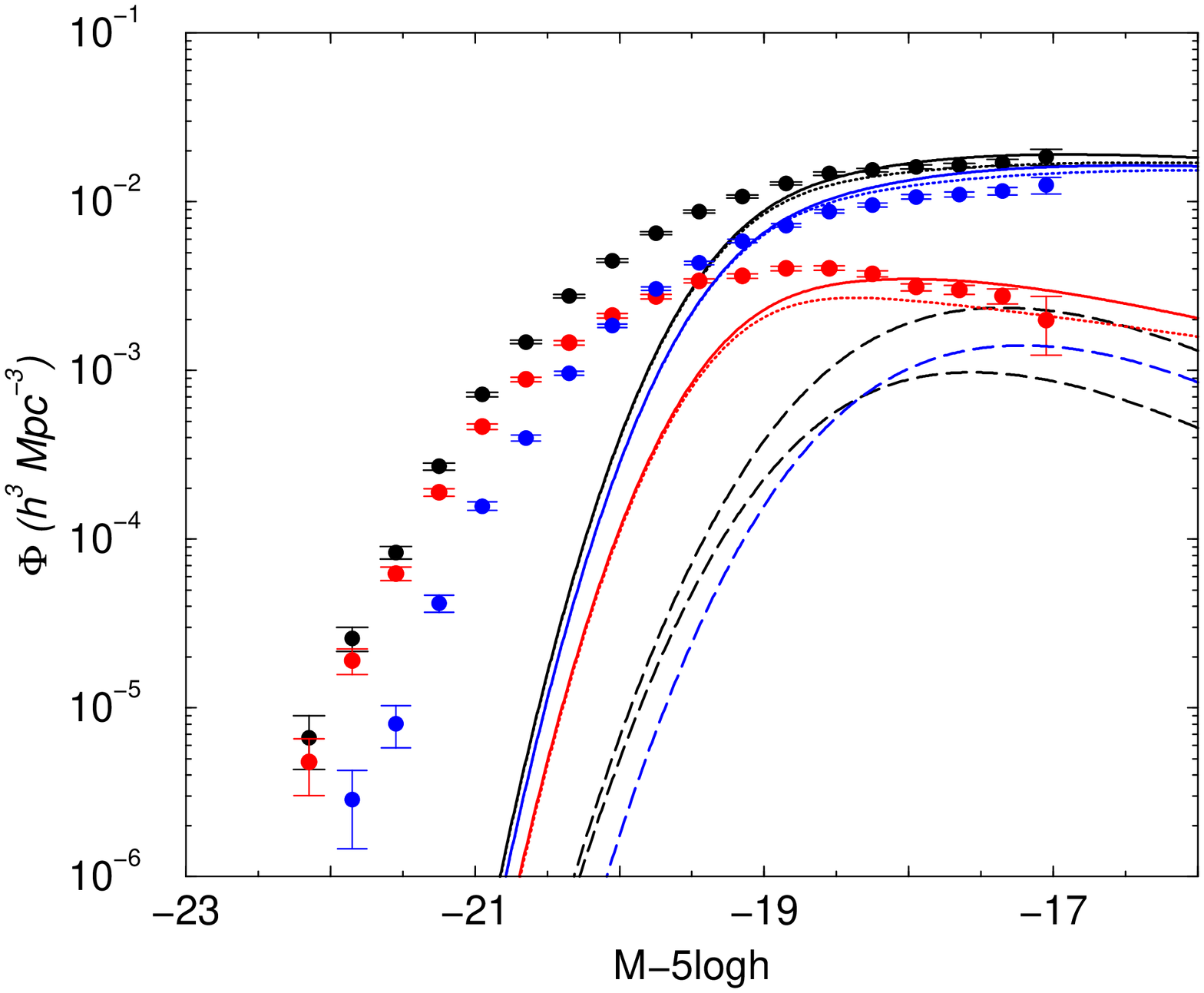,width=\hssize,angle=-90}
\psfig{file=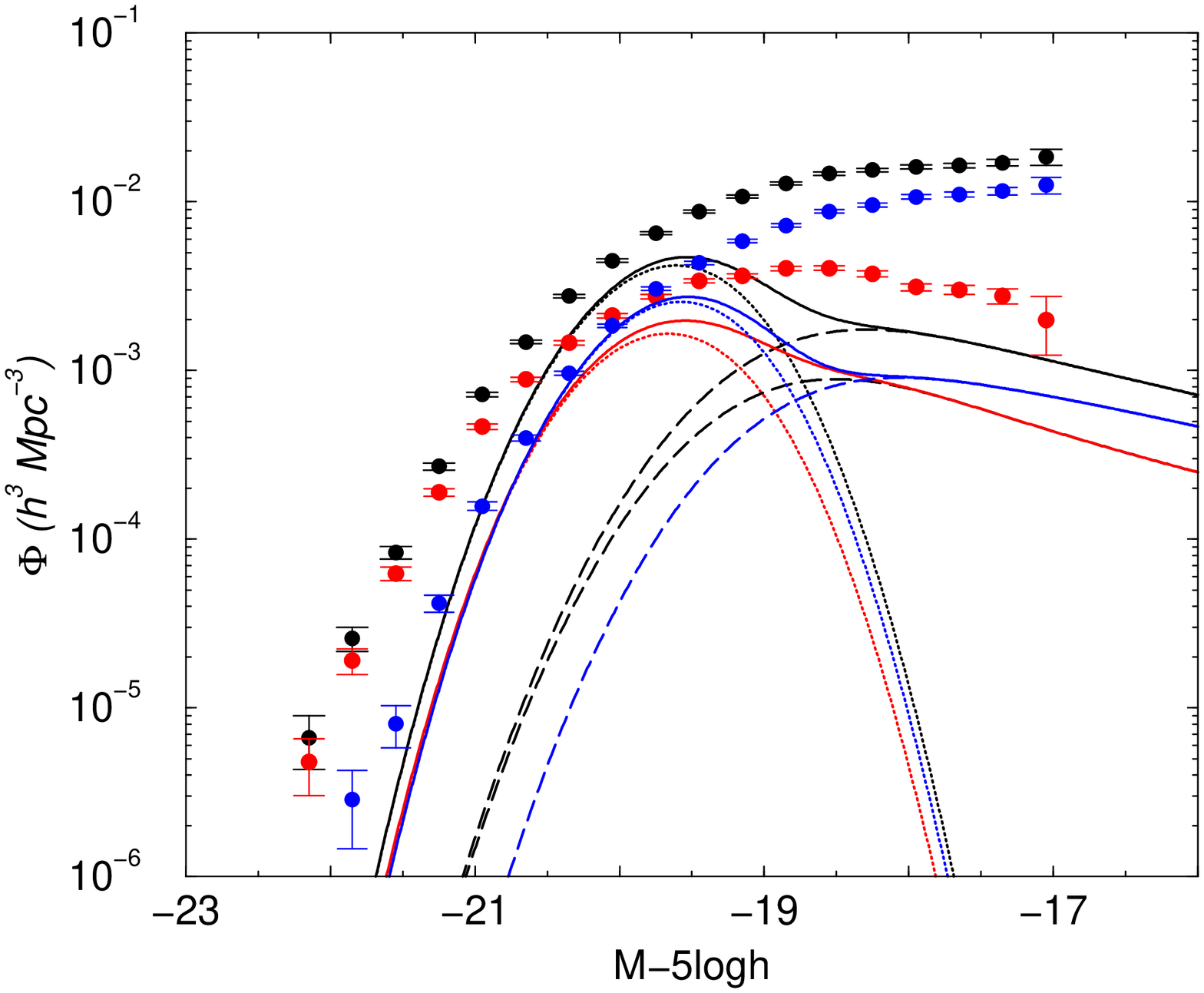,width=\hssize,angle=-90}}
\centerline{
 \psfig{file=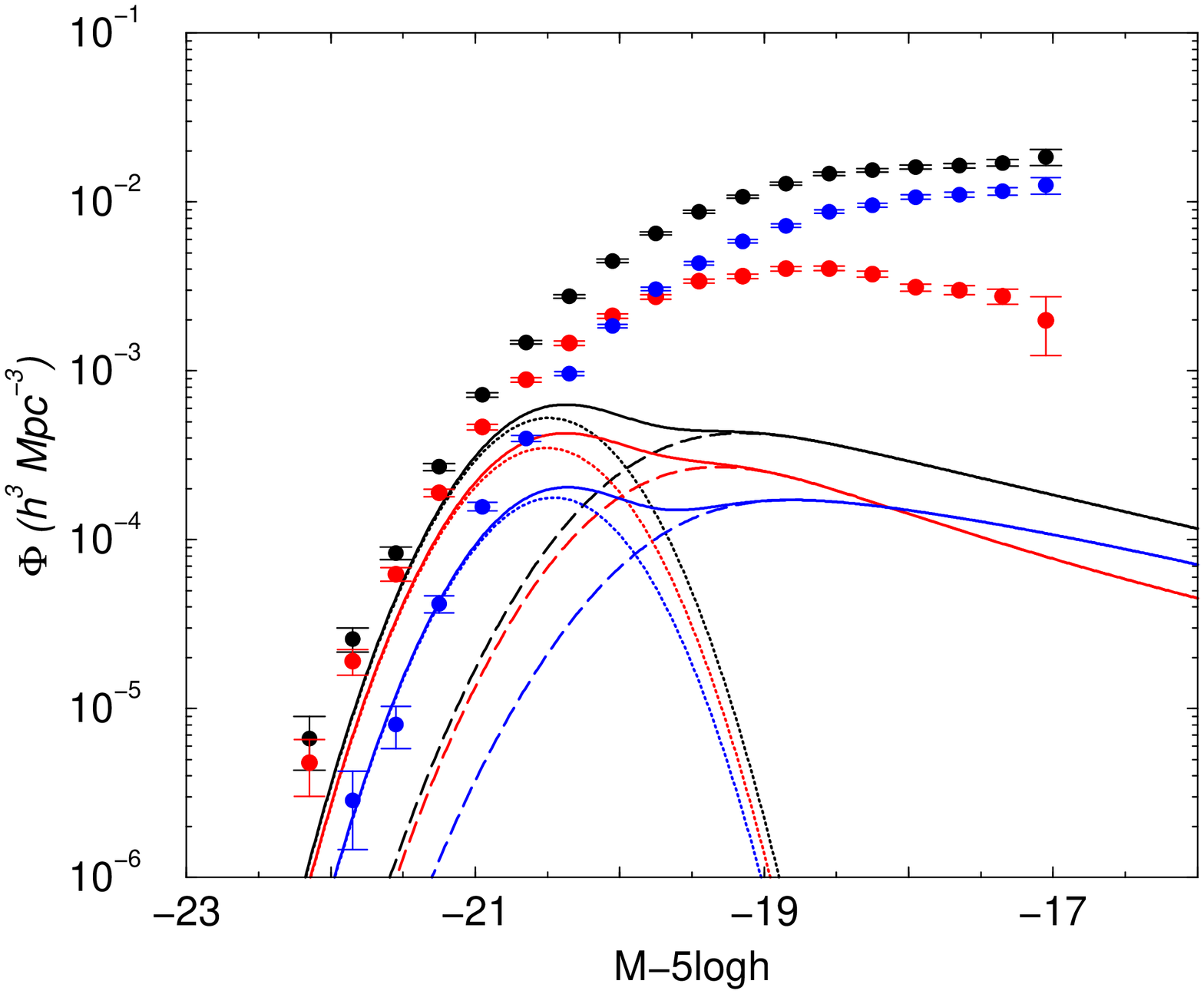,width=\hssize,angle=-90}
\psfig{file=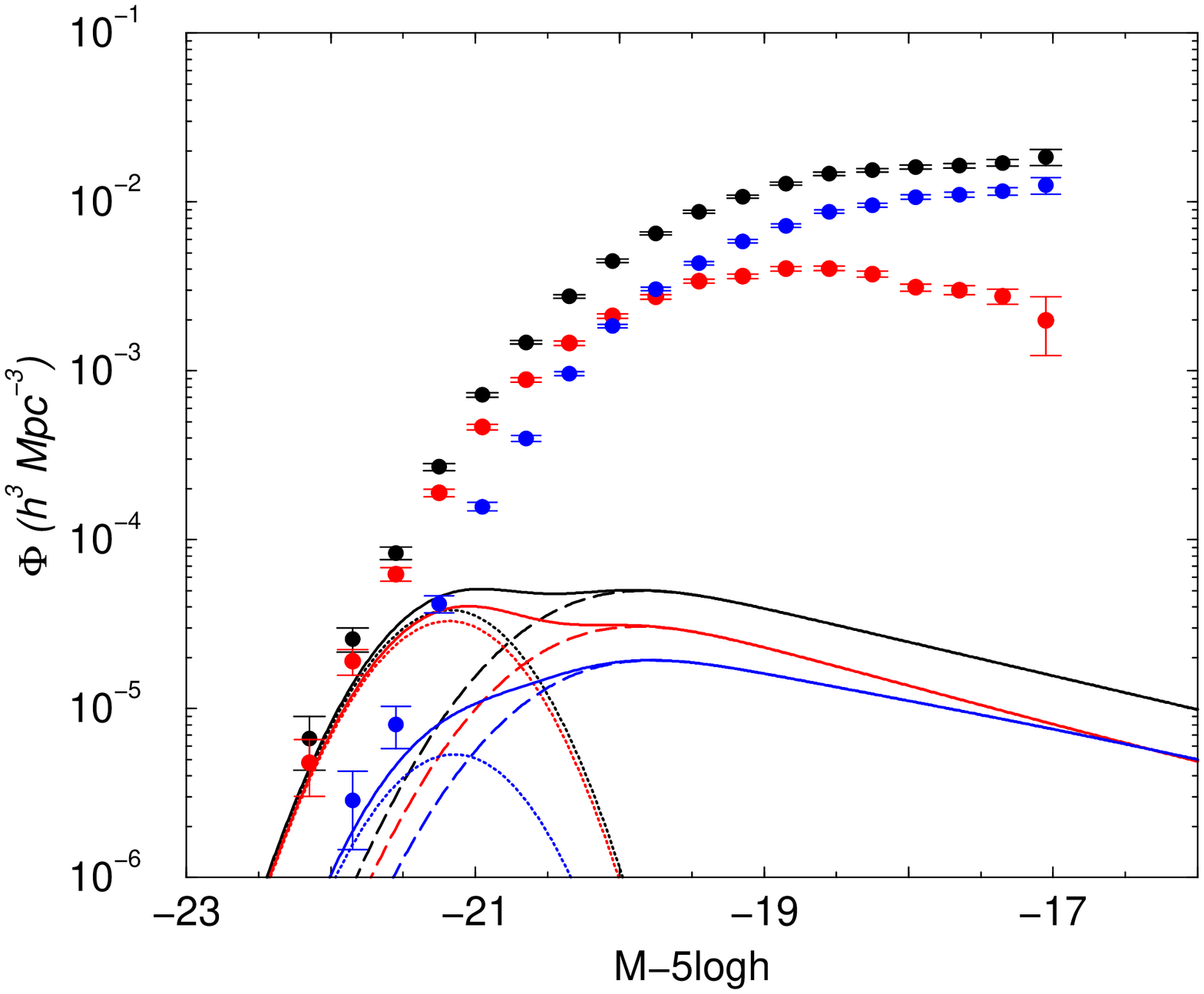,width=\hssize,angle=-90}}
\caption{The mass dependence of the 2dF LF. The plotted data and line styles are same as in Figure~4.
Here, from (a) to (d), we show the contribution to the luminosity
function from a galaxies in the mass ranges $10^9$ M$_{\sun}$ to $10^{12}$  M$_{\sun}$,
$10^{12}$ M$_{\sun}$ to $10^{13}$  M$_{\sun}$,
$10^{13}$ M$_{\sun}$ to $10^{14}$  M$_{\sun}$, and
$10^{14}$ M$_{\sun}$ to $10^{16}$  M$_{\sun}$ respectively. Central galaxies  in low mass halos
determine the shape of the LF at the faint-end, while bright galaxies in massive halos, such as groups and
clusters, determined the LF at the bright-end, where an exponential-like cut off is observed.
}
\end{figure*}

\begin{figure*}
\centerline{\psfig{file=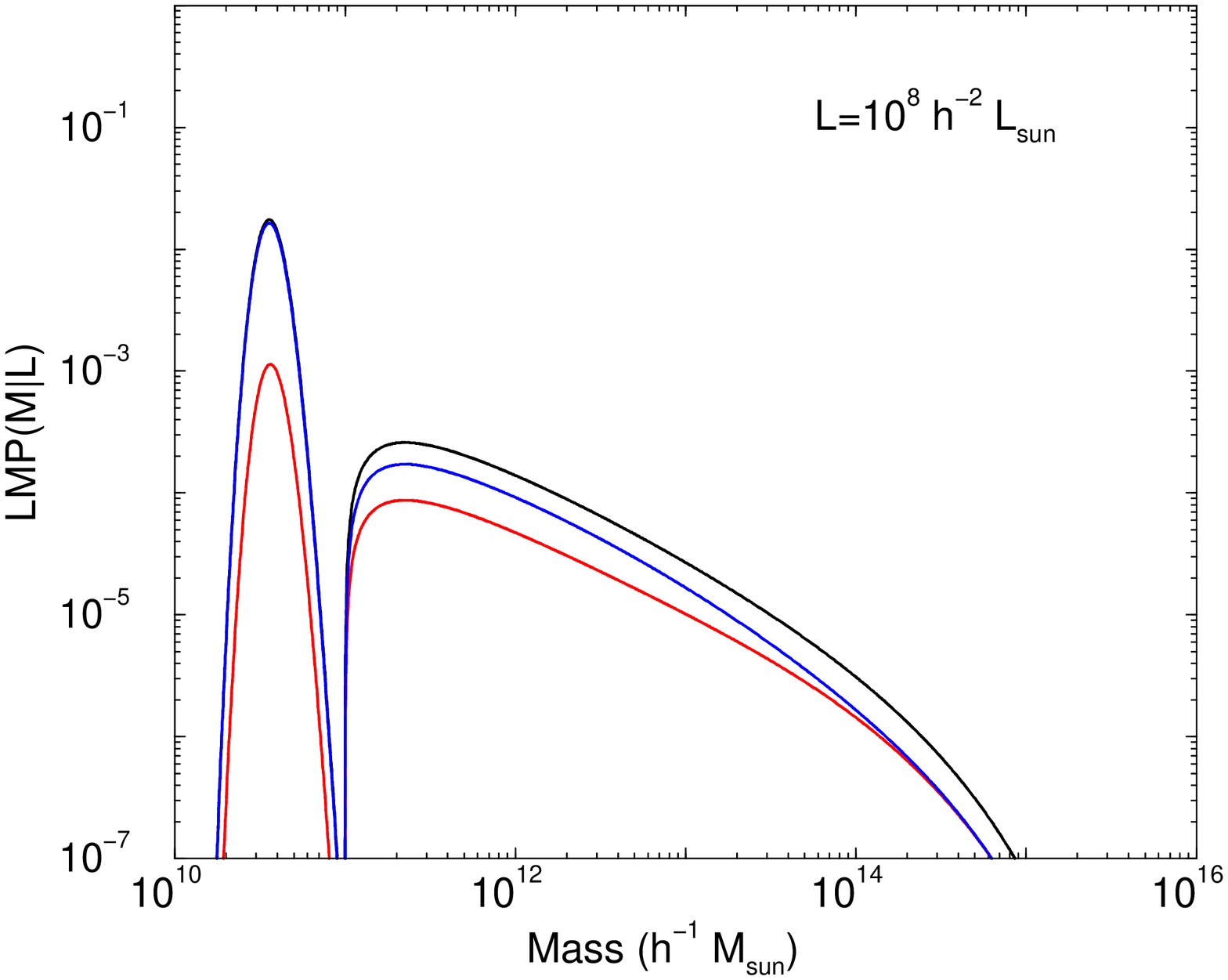,width=\hssize,angle=-90}
\psfig{file=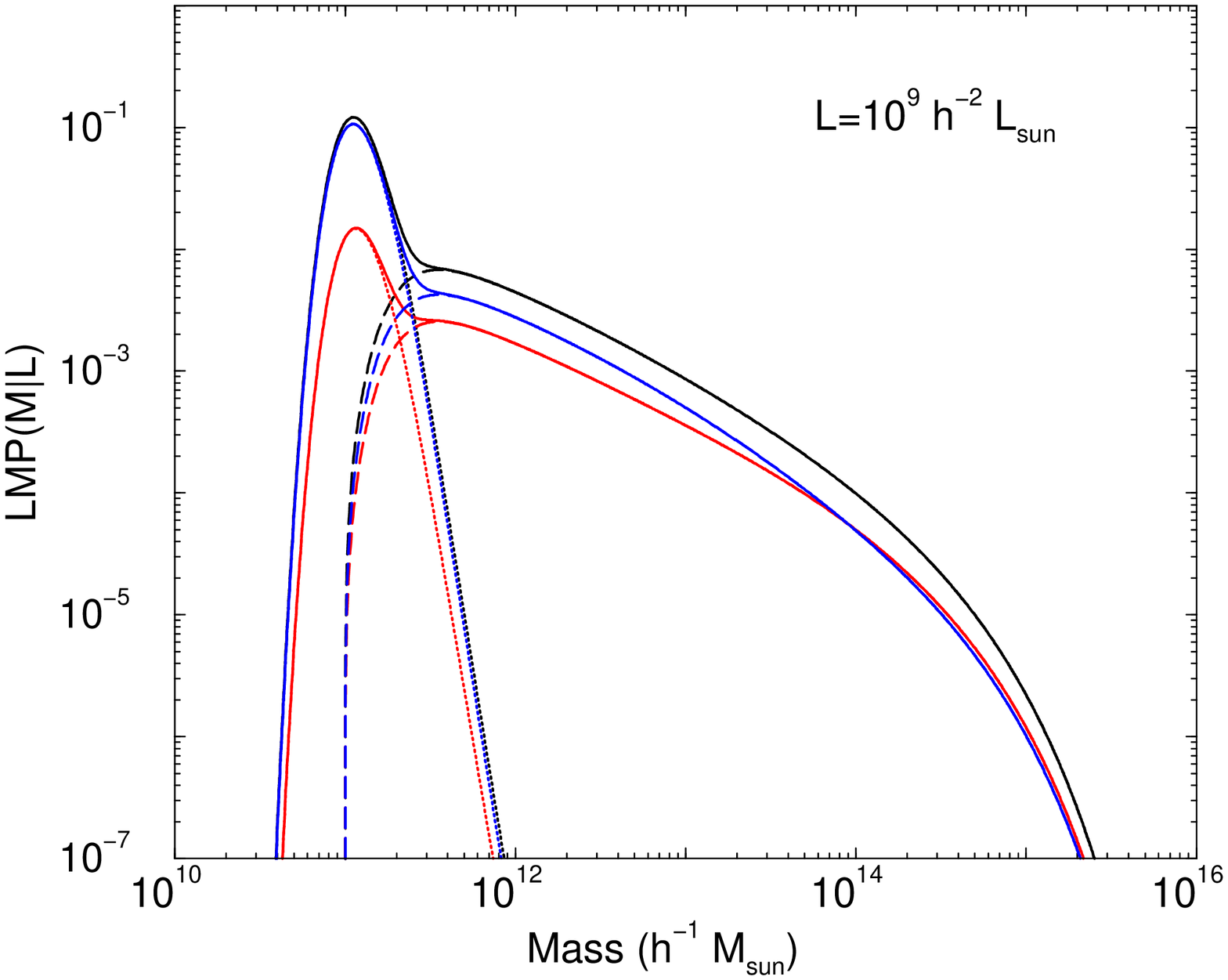,width=\hssize,angle=-90}}
\centerline{\psfig{file=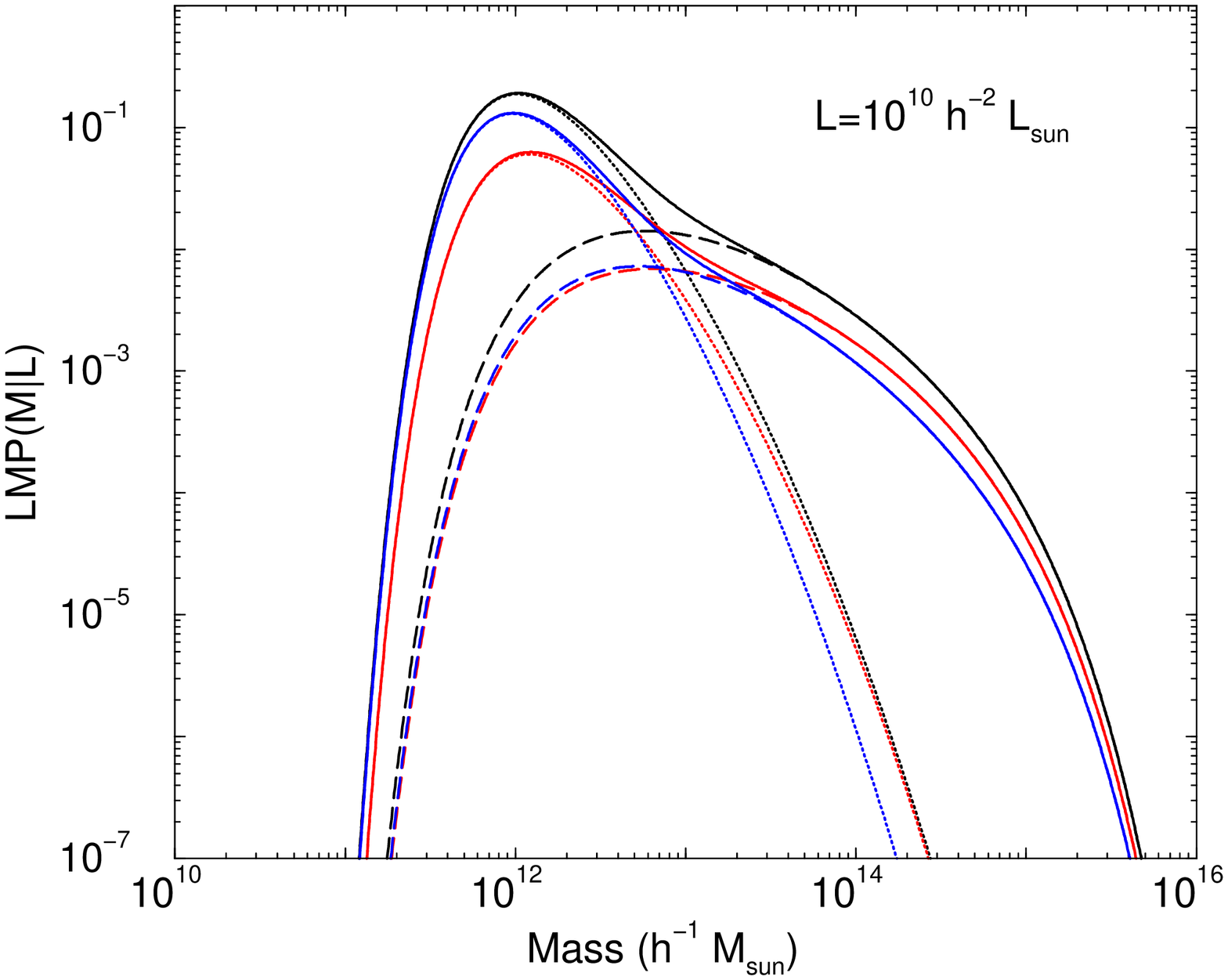,width=\hssize,angle=-90}
\psfig{file=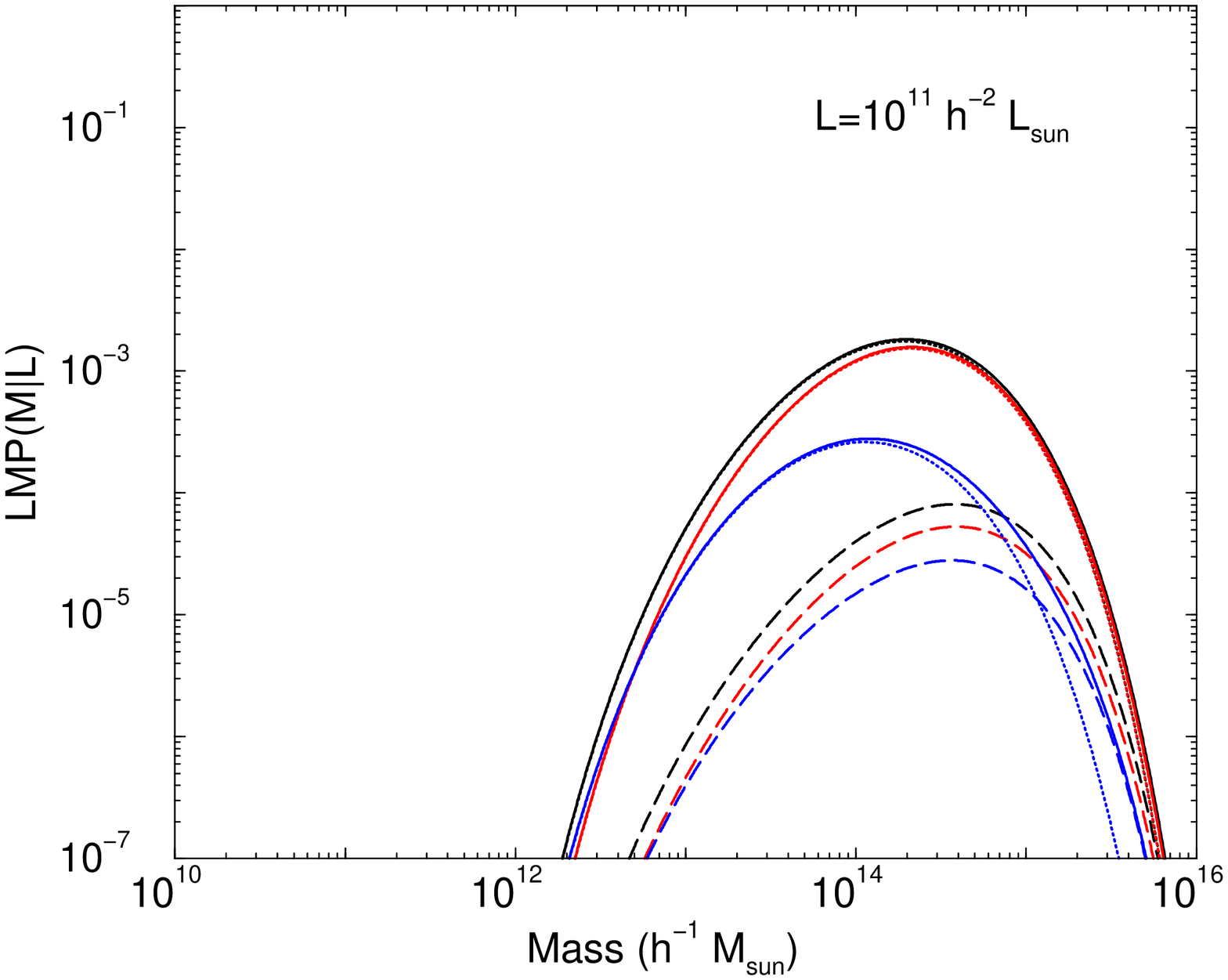,width=\hssize,angle=-90}}
\caption{The conditional probability distribution function of halo mass $P(M|L)$ to host a galaxy of the given
luminosity,  as function of the halo mass.   The black lines are the total galaxy sample, while red and blue lines show the sample divided to
early- and late-type galaxies. These are the same models that were used to describe the LFs of  Croton et al. (2004) and
shown in Figure~4. From (a) to (d), we show  these probabilities
for $L=10^8 h^{-2}$ L$_{\sun}$, $L=10^9 h^{-2}$ L$_{\sun}$, $L=10^{10} h^{-2}$ L$_{\sun}$, and
$L=10^{11} h^{-2}$ L$_{\sun}$, respectively. 
The low mass end peak, which tends to be narrow for lower luminosities,
are the central galaxies, while the tail to high masses, is associated with satellite galaxies.
The width of the central galaxy luminosity peak increases since the $L_c(M)$ relation increases
while  flattening such that one encounters a fractionally higher mass range with an increase in
luminosity. Low luminous late-type galaxies are found in low-mass halos, while luminous red galaxies 
are in halos with masses corresponding to galaxy groups and clusters.}
\end{figure*}

\begin{figure}
\centerline{\psfig{file=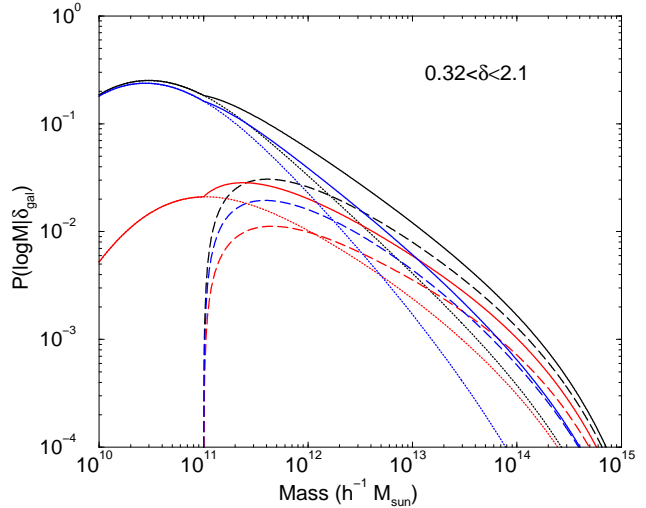,width=\hssize,angle=-90}}
\caption{The probability distribution function of halo masses hosting 2dF galaxies, when averaged over the LF.
The black lines are the total galaxy sample, while red and blue lines show the sample divided to
early- and late-type galaxies following measurements by Croton et al. (2004).}
\end{figure}

\section{Luminosity Functions}

Given our model for the CLFs, we can now construct the LF, which is an average of CLFs in mass with
the halo mass distribution given by the mass function. Here, we use the
Sheth \& Tormen (1999; ST) mass function $dn/dM$ for dark matter 
halos. This mass function is in better agreement with numerical simulations (Jenkins et al. 2001), when compared to
the more familiar Press-Schechter (PS; Press \& Schechter 1974) mass function.

Given the mass function, the galaxy LF is 
\begin{equation}
\Phi_i(L) = \int_0^\infty \Phi_i(L|M) \frac{dn}{dM} dM \,,
\end{equation}
where $i$ is an index for early and late type galaxies.
The conditional luminosity function for each type involves the sum of central and satellites.
To understand our model for the LF, we will plot these two divisions, as well as the sum, separately in each of
the LF figures shown in this paper.

\begin{figure*}
\centerline{\psfig{file=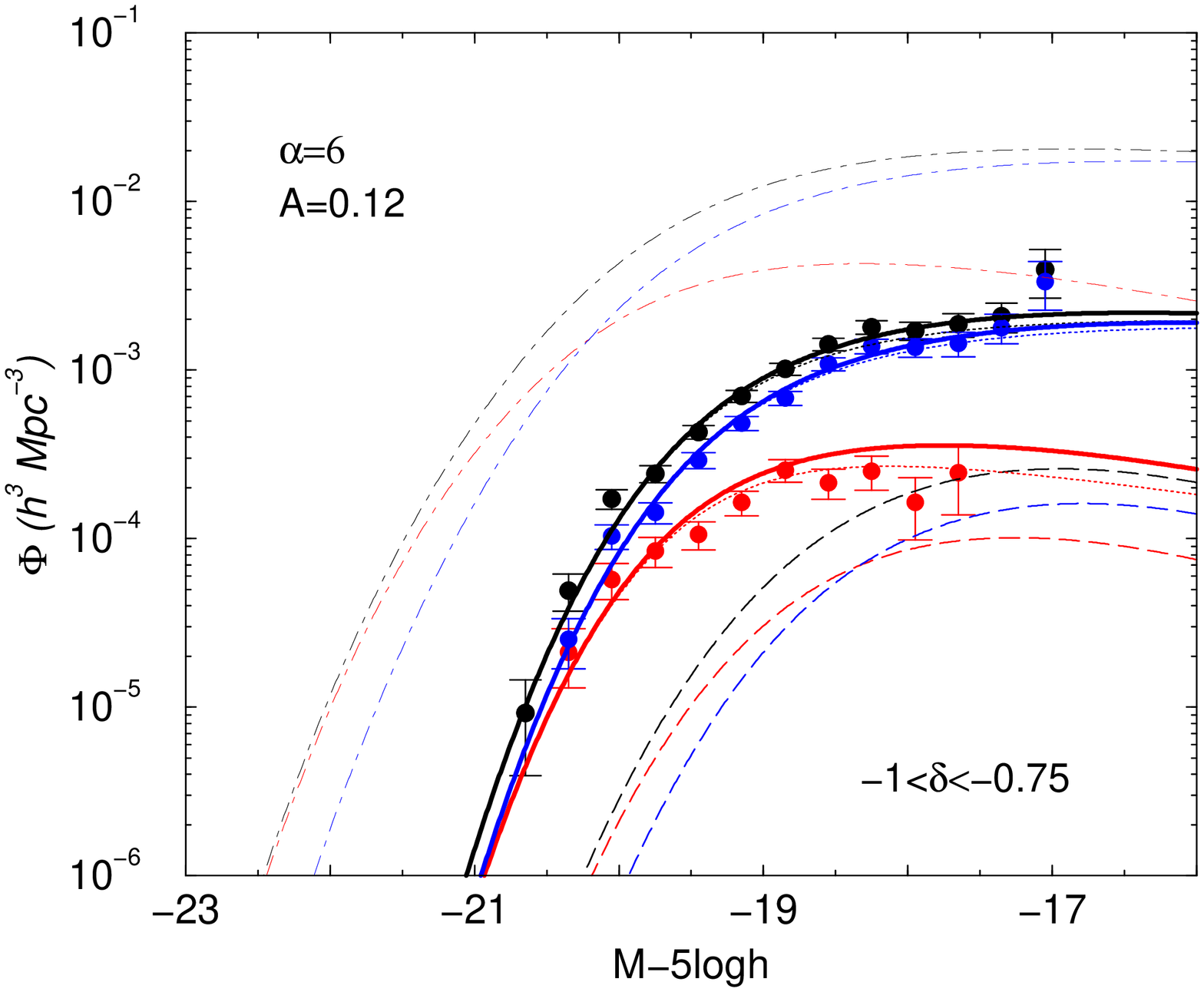,width=\hssize,angle=-90}
\psfig{file=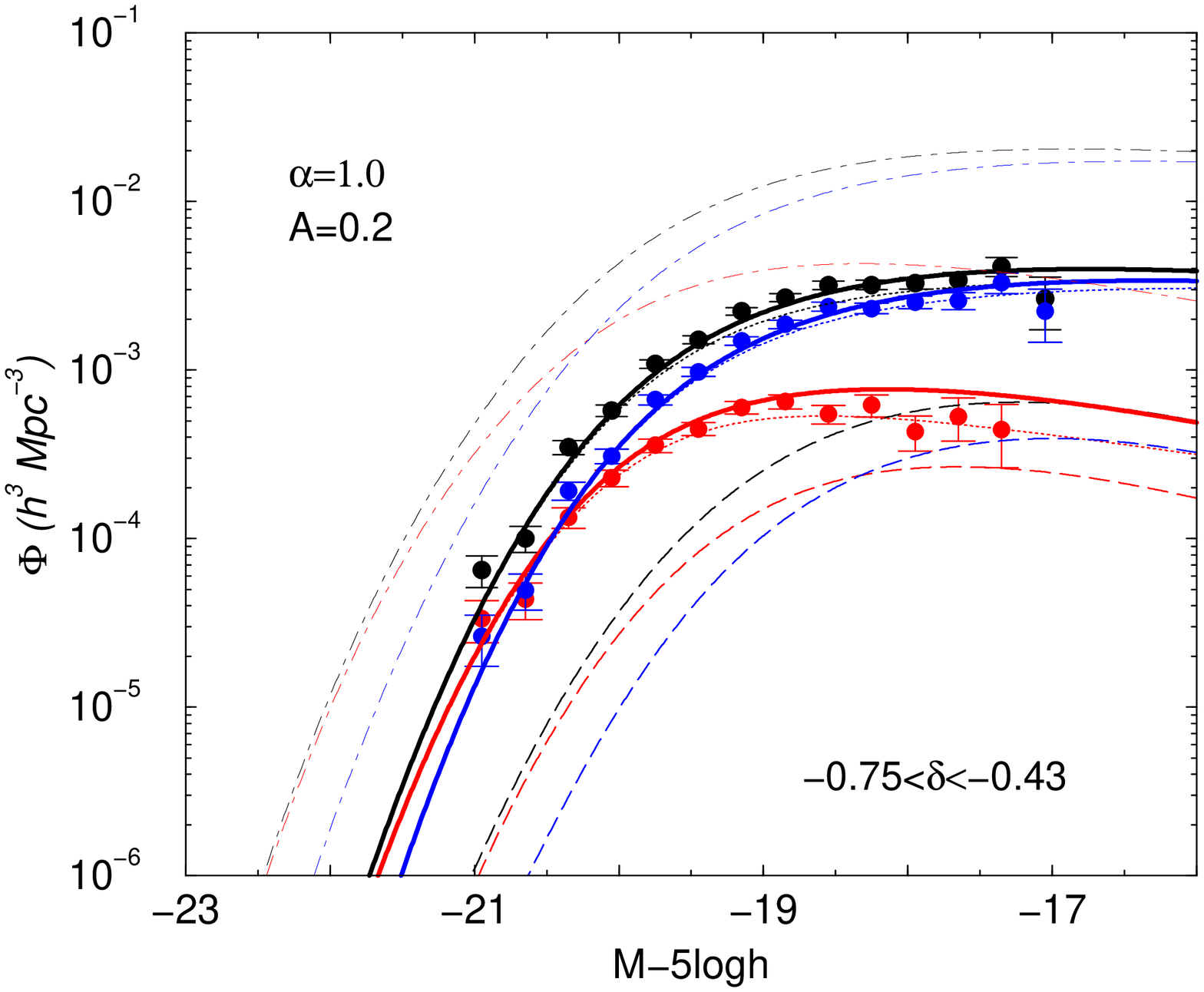,width=\hssize,angle=-90}}
\centerline{\psfig{file=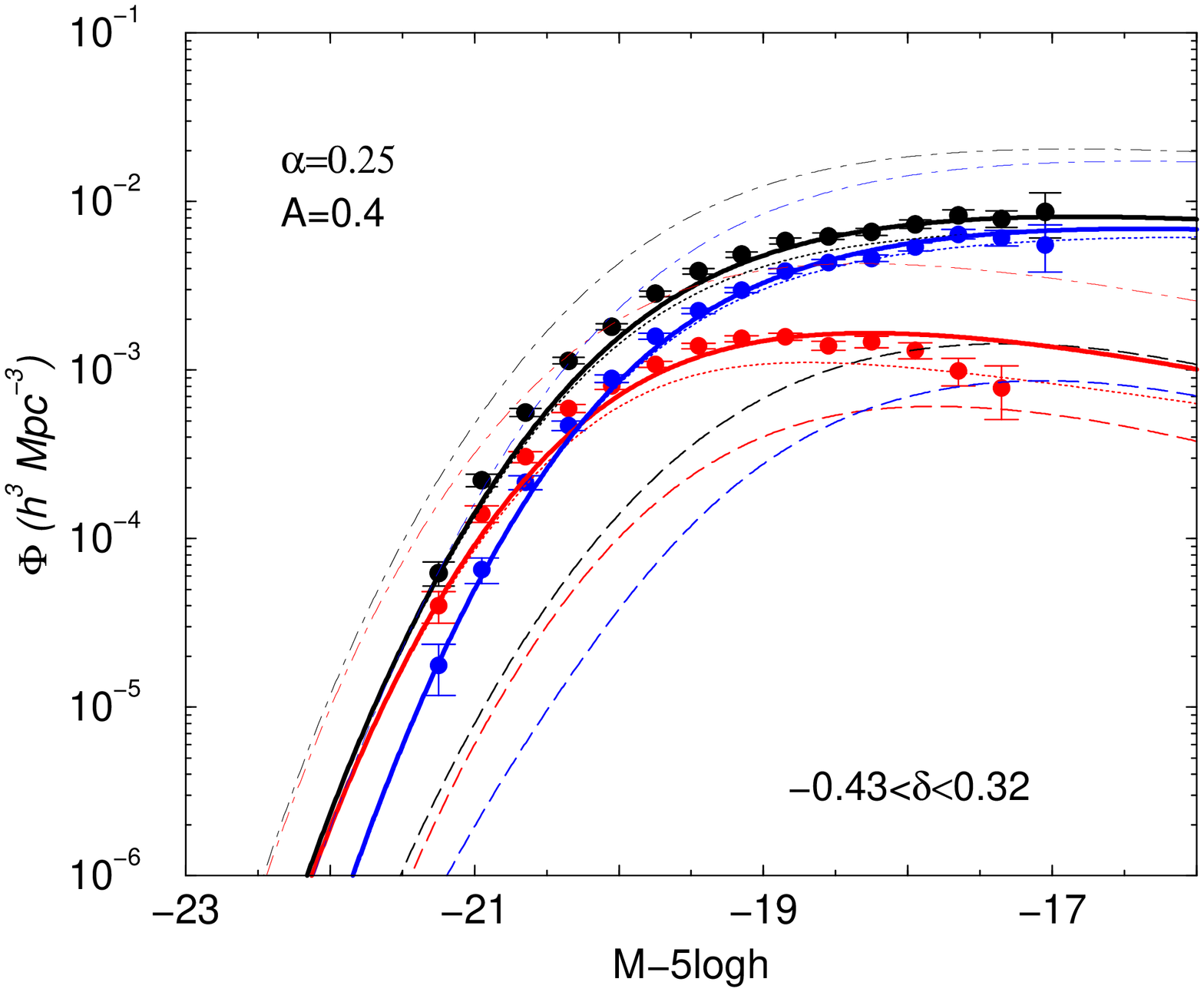,width=\hssize,angle=-90}
\psfig{file=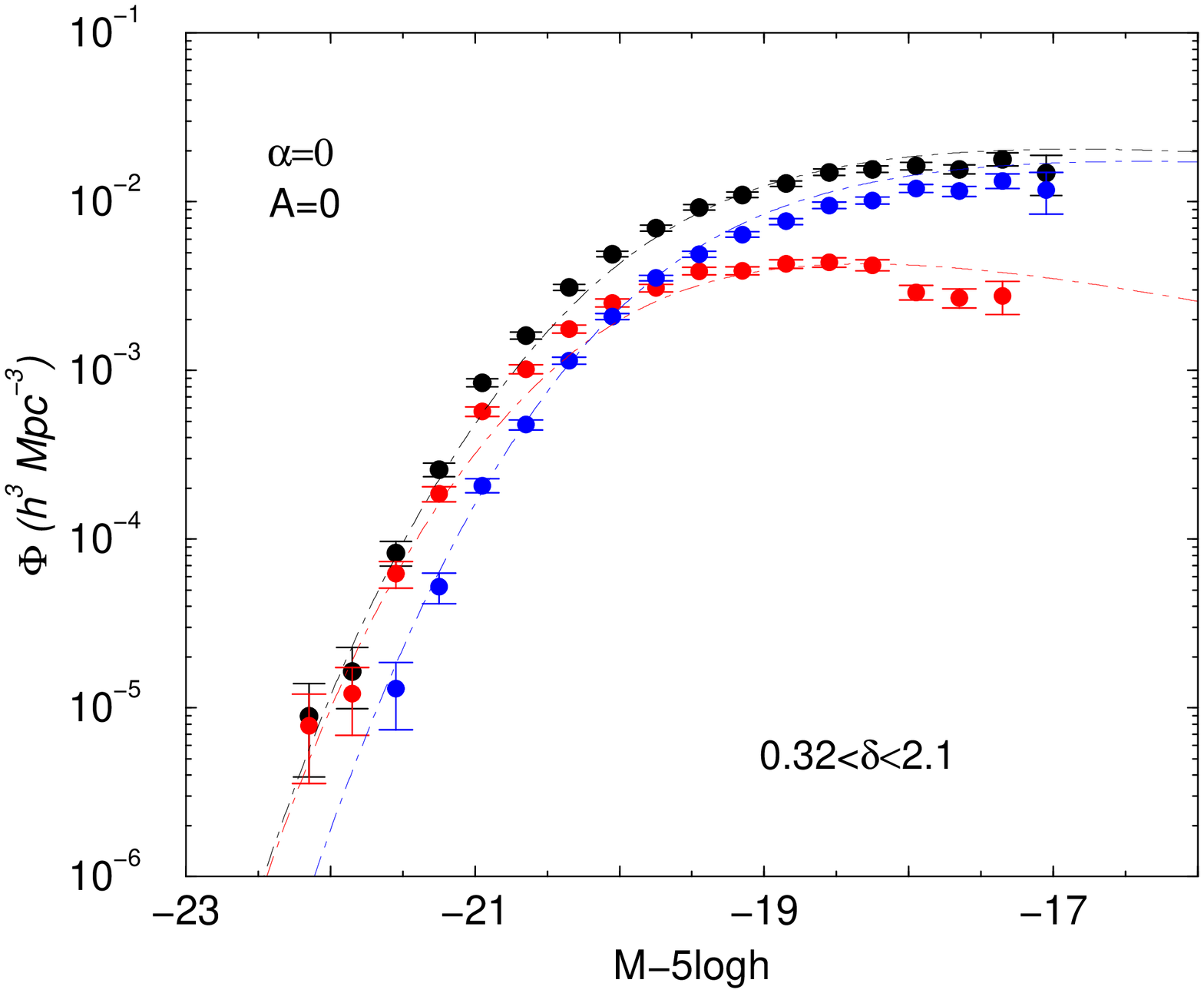,width=\hssize,angle=-90}}
\centerline{\psfig{file=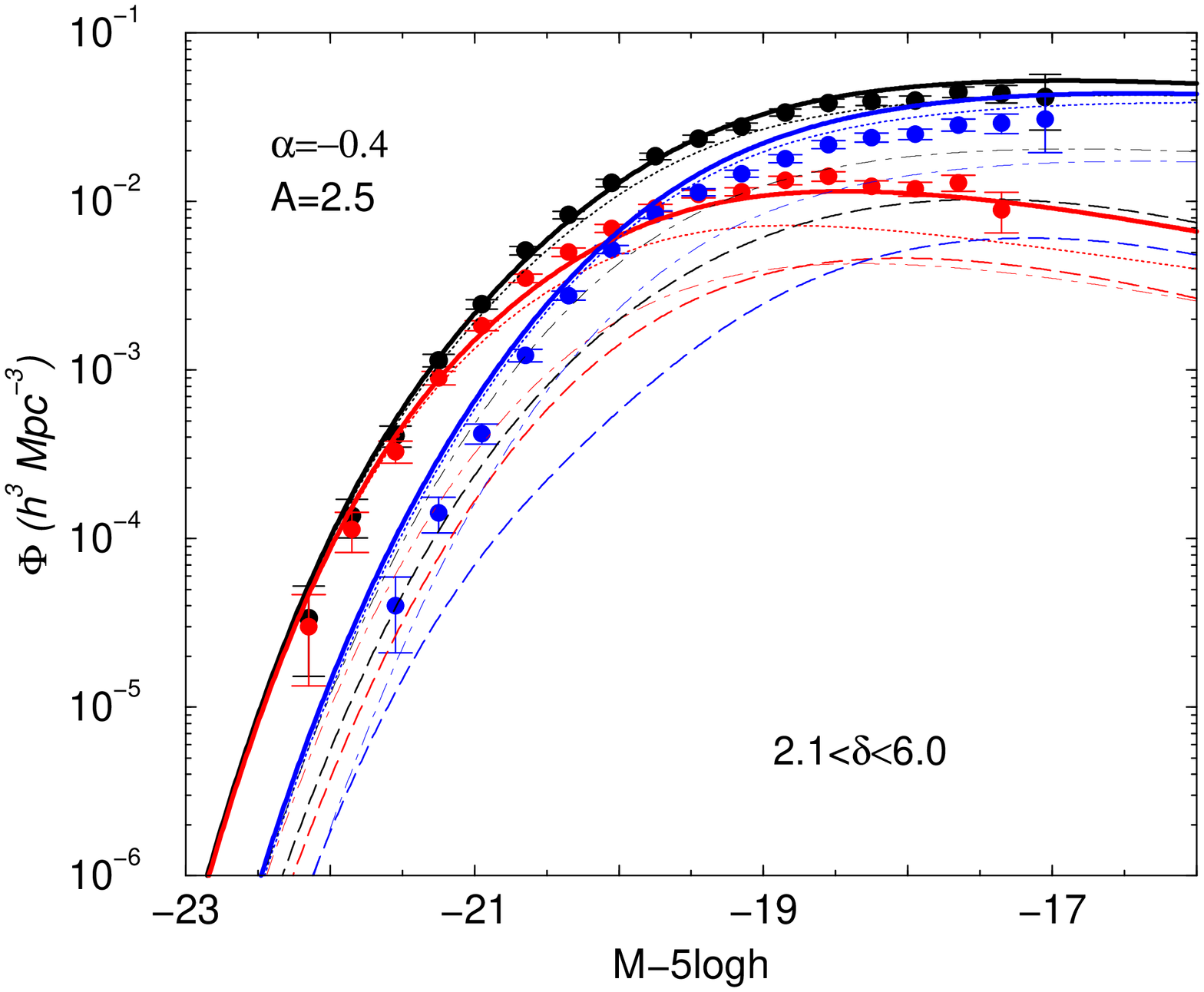,width=\hssize,angle=-90}
\psfig{file=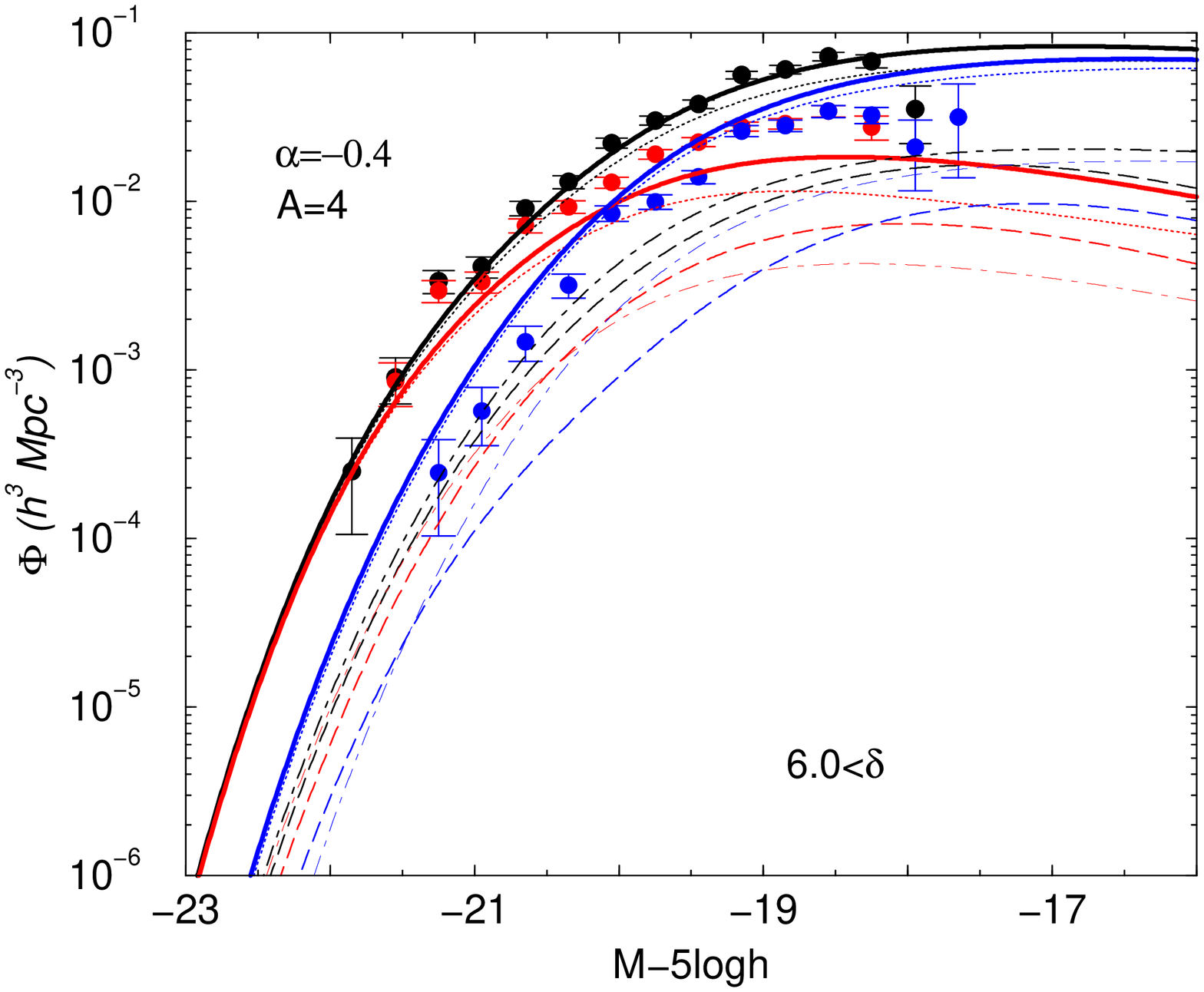,width=\hssize,angle=-90}}
\caption{The environmental LFs, or $\Phi(L|\delta_{\rm gal})$, where $\delta_{\rm gal}$ is the
galaxy overdensity measured over a volume of 8 $h^{-1}$ Mpc. The measured data are from
Croton et al. (2004). From (a) to (f), the overdensity varies from $\sim -0.9$ to greater than 6.
In each of the panels, we show the ``best-fit'' models for the conditional mass function,
$dn(M|\delta_{\rm gal})/dM$, based on parameters $A(\delta_{\rm gal})$ and $\alpha(\delta_{\rm gal})$
following Eq.~\ref{eqn:dndmdelta}.
The line styles are same as Figure~4. For comparison, we also show the average LF in Figure~4(b),
with dot-dashed lines. }
\end{figure*}

\begin{figure*}
\centerline{\psfig{file=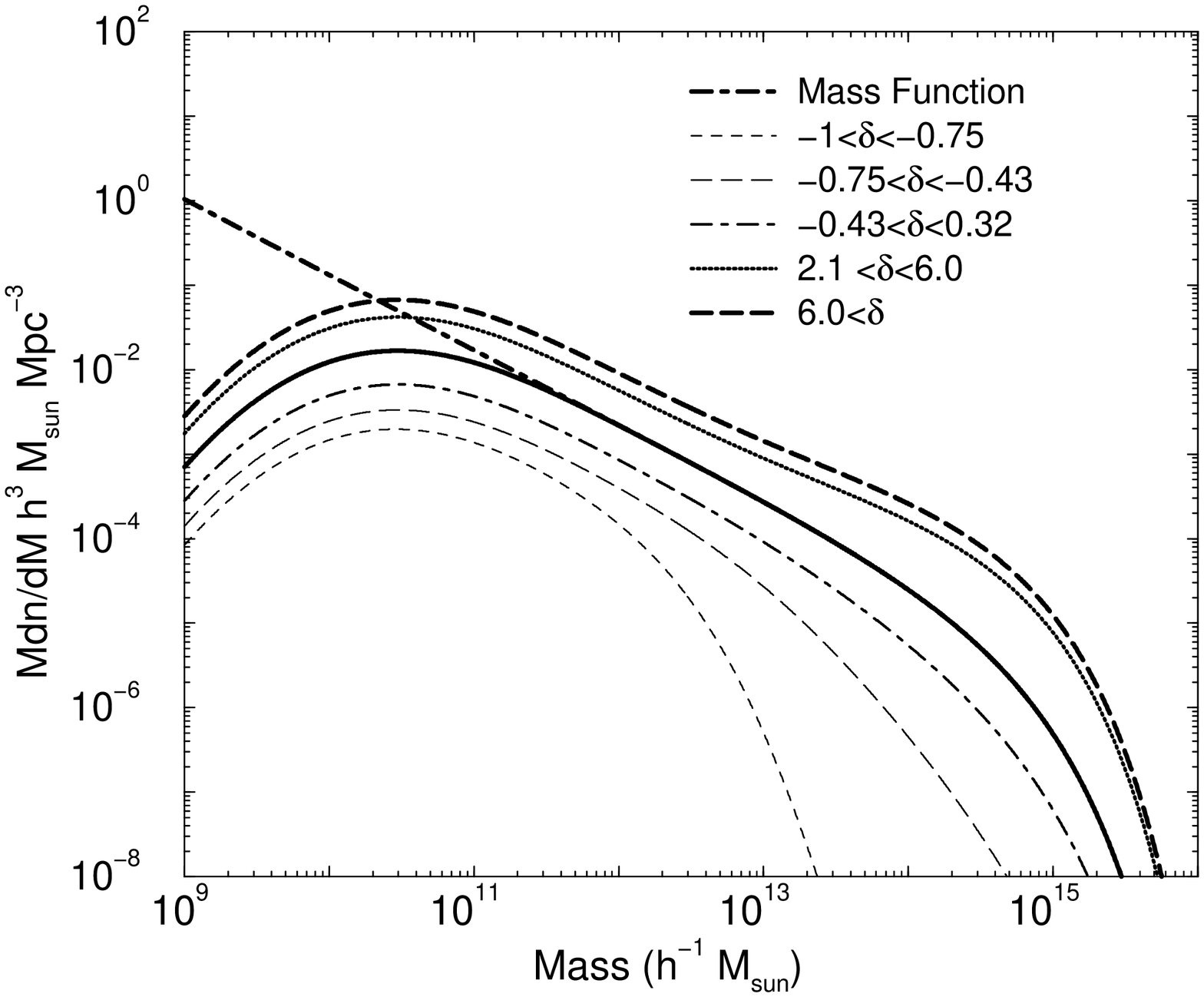,width=\hssize,angle=-90}
\psfig{file=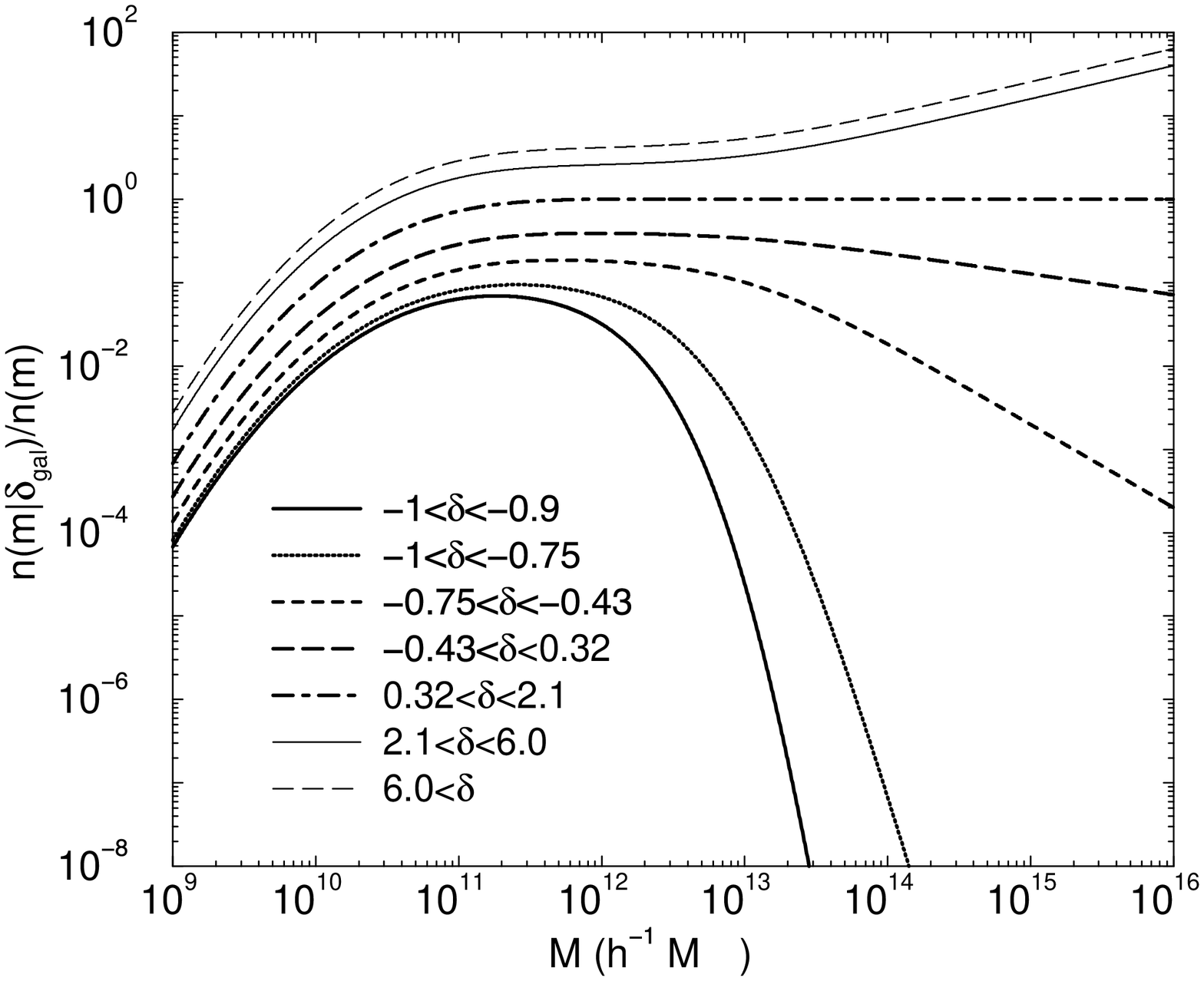,width=\hssize,angle=-90}}
\caption{(a) The predicted mass functions, as a function of the environment captured by the galaxy overdensity
over a size scale of 8 $h^{-1}$ Mpc. 
The dot-dashed line is the standard ST mass function, while other curves show the mass
function from voids with $\delta_{\rm gal} < 0$ to dense regions with $\delta > 1$. (b) The ratio of
environmental mass functions to the mean mass function. Note the sharp cut-off  at high masses when $\delta < -0.5$.
In the case of dense regions when $\delta \gg 1$, the mass function increase at all mass scales, relative to the mean,
but with a greater increase at high masses.
}
\end{figure*}

\begin{figure*}
\centerline{\psfig{file=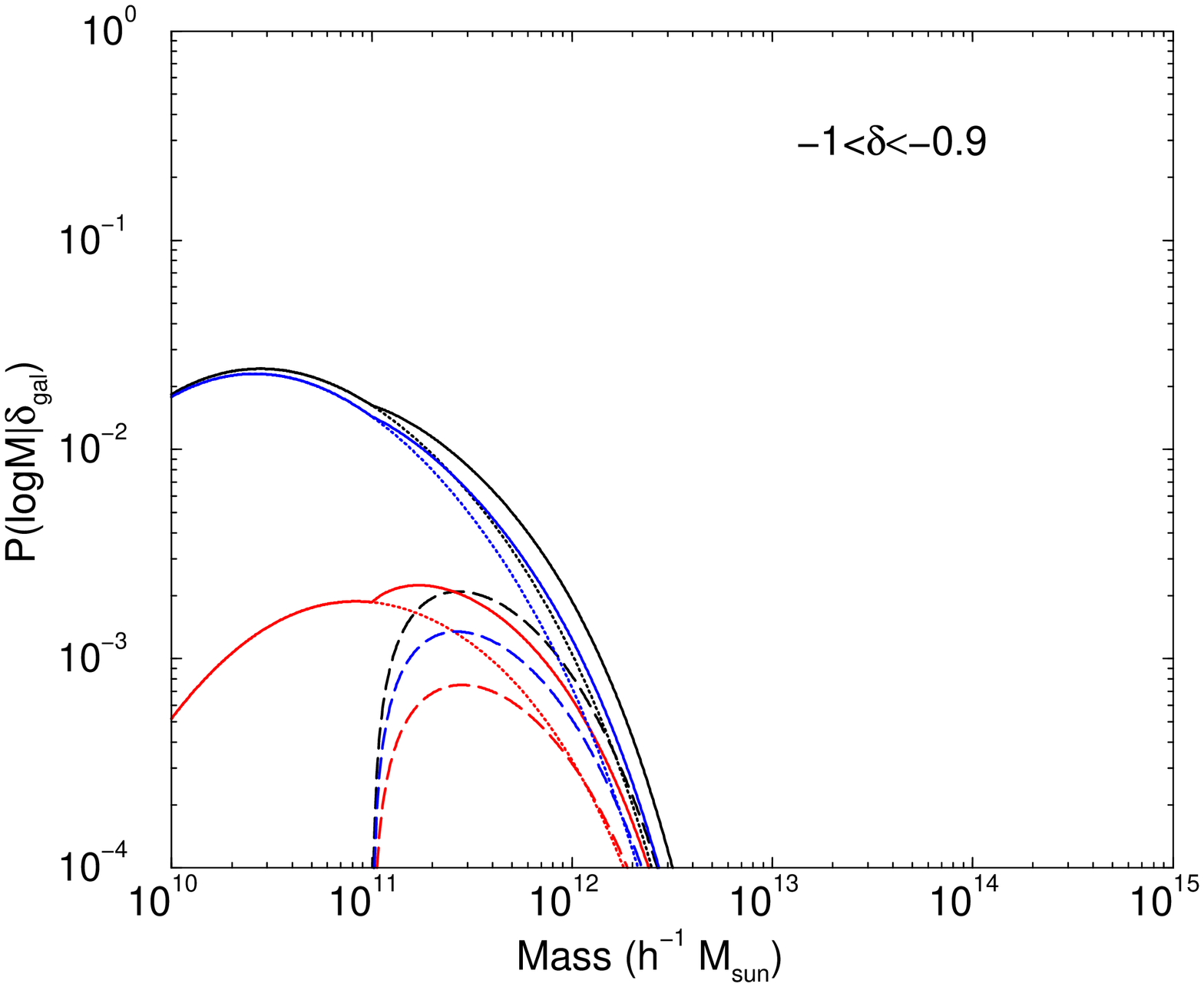,width=\hssize,angle=-90}
\psfig{file=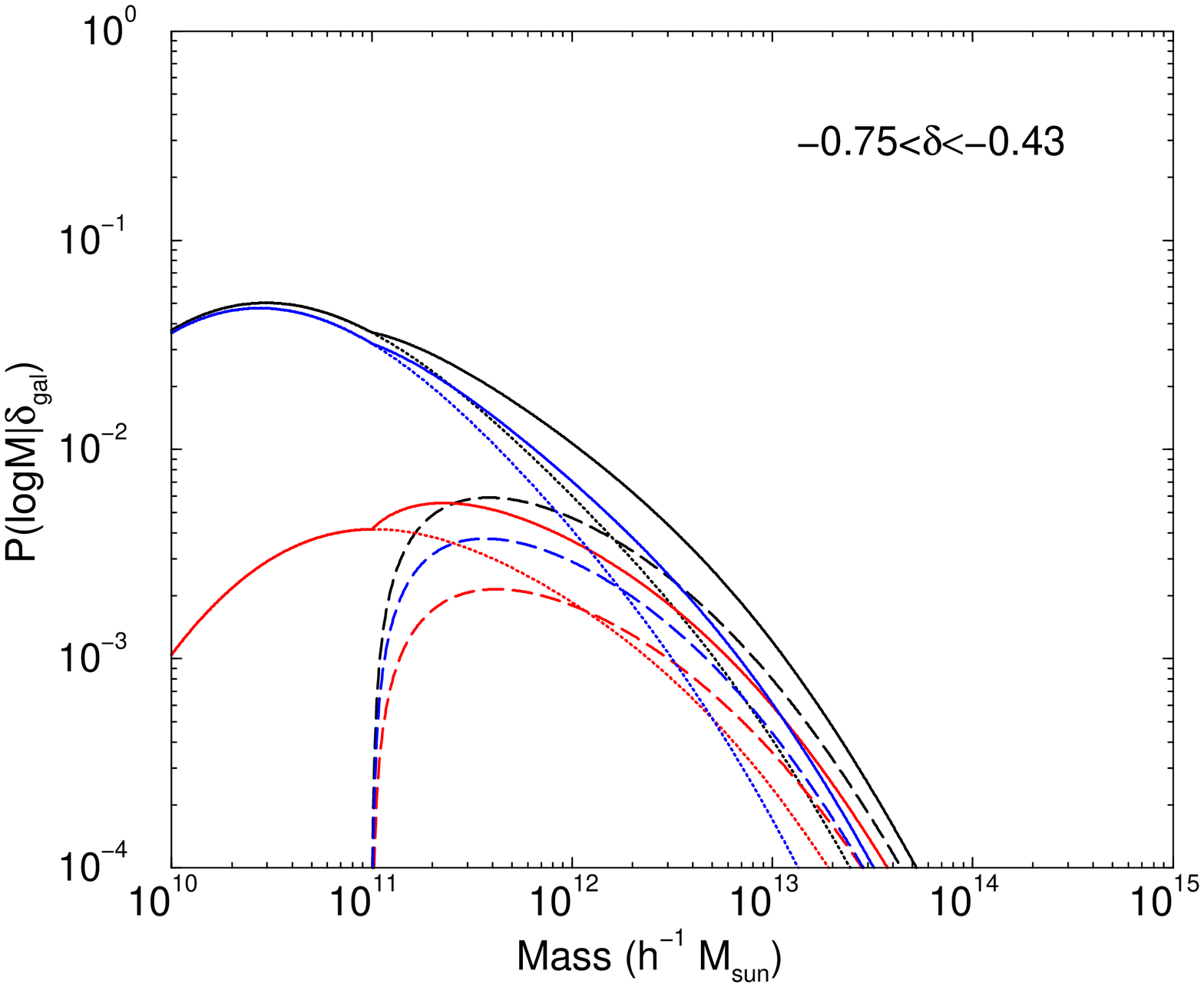,width=\hssize,angle=-90}}
\centerline{\psfig{file=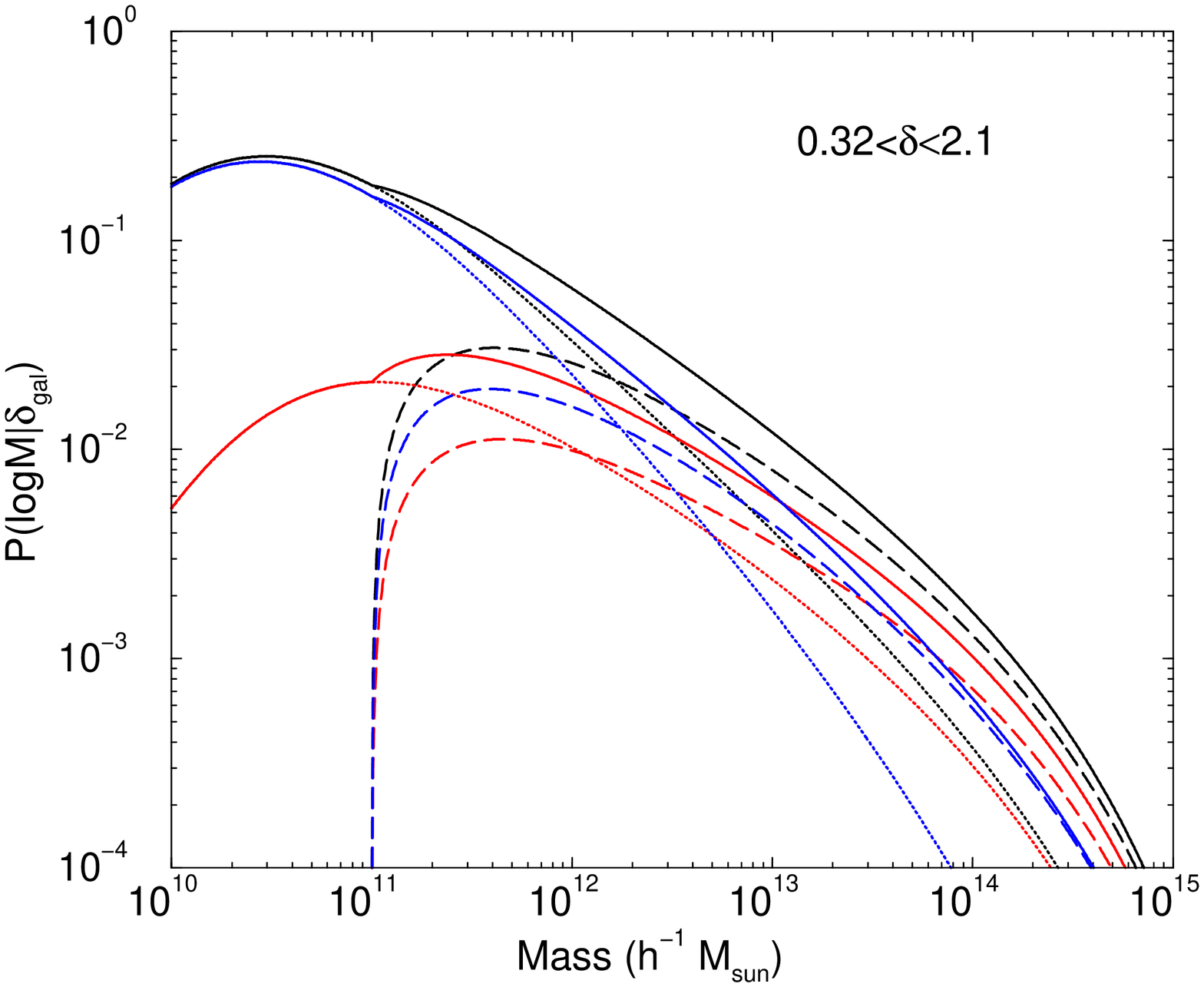,width=\hssize,angle=-90}
\psfig{file=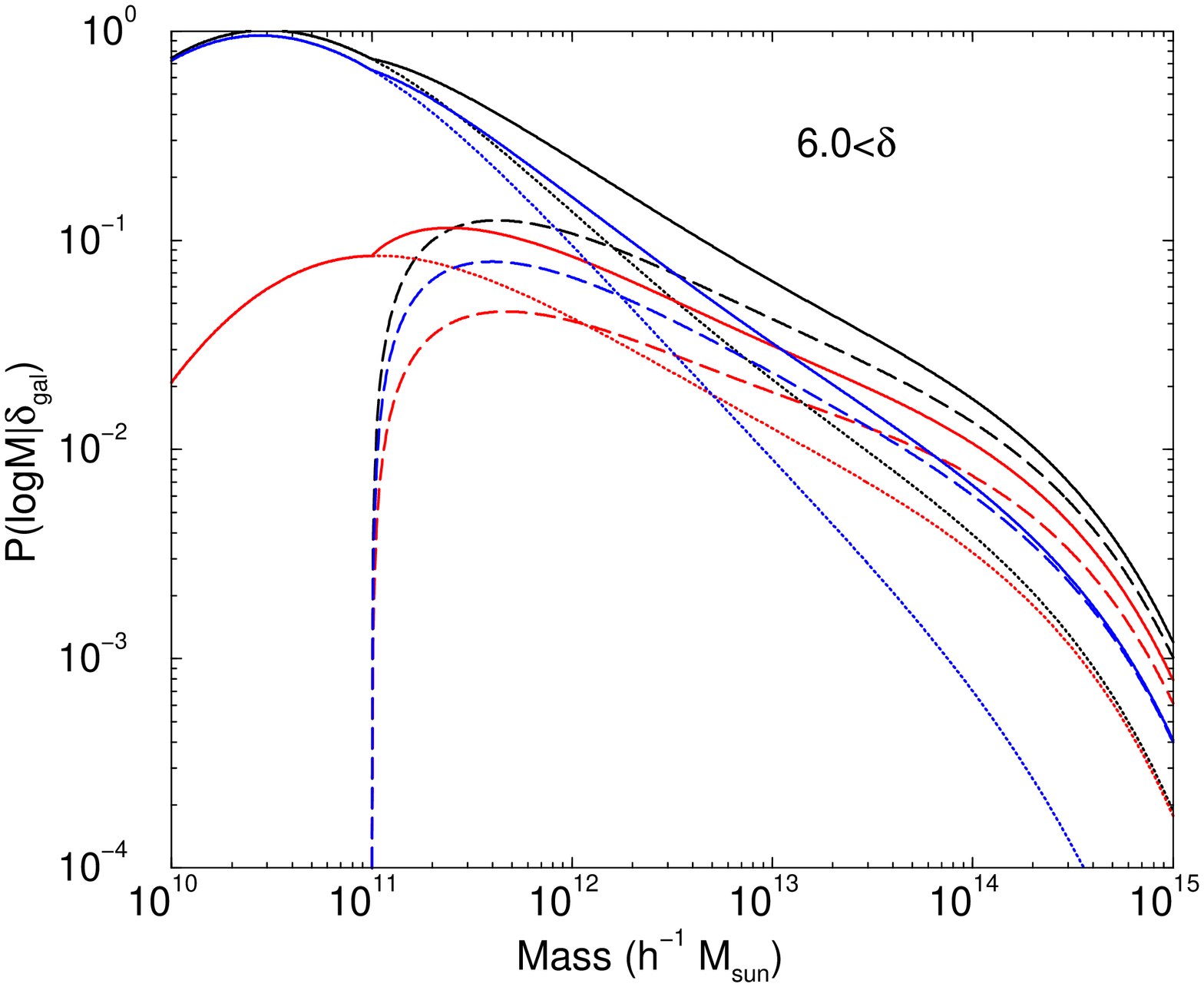,width=\hssize,angle=-90}}
\caption{The probability distribution 
function of halo masses hosting 2dF galaxies as a function of the environment as measured by the
galaxy overdensity in 8 $h^{-1}$ Mpc volumes. Here, we plot $P(\log M |\delta_{\rm gal})$.
The black lines are the total galaxy sample, while red and blue lines show the sample divided to
early- and late-type galaxies following measurements by Croton et al. (2004).
Form (a) to (d), we show four ranges in $\delta_{\rm gal}$
corresponding to both under dense and over dense regions. In the case of under dense regions, halos
with mass $10^{11}$ M$_{\sun}$ host most galaxies and the the fraction of galaxies
is dominated by the late types (blue galaxies). In the case of over dense regions,
 a variety of mass scales contribute to the luminosity function of that region,
with early-types dominating the fraction from massive halos, while late-types dominate
the fractional contribution from low mass halos.}
\end{figure*}

Figure~4 presents a general fit to the 2dF LF, where in Fig.~4(a), we assume $f_c(M)=1$. While the model
describes data adequately at magnitudes below -19, the fit at the low-end of luminosity is poor.
Our model suggests that the  low-end slope of the LF should be around $-1.33$ to $-1.25$, with the
latter coming from our assumption that $L_c(M) \sim M^4$ at low mass halos (Cooray \& Milosavljevi\'c 2005b).
The flattening of the slope at the low-end of luminosity, as measured in the 2dF LF, may be
a reflection of either (1) galaxy selection function such that faint galaxies are missed, or (2)
a real effect in the Universe such that low mass halos do not host a large number of central galaxies.
The difference from the model in Fig.~4(a), when compared to data, can be reduced with the
inclusion of a mass-dependent $f_c(M)$ function. In Fig.~4(b), we show the case with equation~\ref{eqn:fcm}.
Note that the underlying reasons for this mass function can either be a selection effect or a real effect.
We cannot distinguish between the two possibilities; but, it is likely that most selection biases are
already accounted when constructing the LF and the effect we are seeing mostly is due to the fact that
not all low mass dark matter halos host a galaxy. 

To understand the mass dependence of the LF, in Figure~5, we plot the LF separated in to mass bins
between $10^9$ M$_{\sun}$ to $10^{16}$ M$_{\sun}$. Galaxies in low mass halos dominate the
statistics of the LF at the faint-end; in fact, 2dF luminosity function  at magnitudes fainter than -19
is associated with galaxies in halos with masses below 10$^{12}$ M$_{\sun}$. On the other hand,
the exponential decrease in the Schechter form for the LF at the bright end is associated with
galaxies in halos of mass $10^{13}$ M$_{\sun}$ and above. As shown in Figure~1, the $L_c(M)$ relation begins
to flatten at halo masses above $10^{13}$ M$_{\sun}$. The characteristic luminosity, $L_\star$, can
be identified with the luminosity of galaxies in halos of this mass scale. The exponential drop-off,
instead of a sharp-cut off is a reflection of the scatter in the $L_(M)$ relation above this mass scale
as was explained in Cooray \& Milosavljevi\'c (2005b). To fit the 2dF LF, we require $\Sigma \sim 0.17$;
this is lower than the value of $\sim 0.23$ found in Cooray \& Milosavljevi\'c (2005b) when modeling the
K-band LF of Huang et al. (2003).

To understand the relative distribution of mass given the galaxy luminosity, 
we also calculate the conditional probability distribution $P(M|L)$ that a galaxy of given luminosity $L$ is in
a halo of mass $M$ as (Yang et al. 2003b):
\begin{equation}
P(M|L)\; dM = \frac{\Phi(L|M)}{\Phi(L)} \frac{dn}{dM} dM  \, .
\end{equation}
In Figure~6, we show these conditional probability distribution  functions, as a function of the halo mass,
when $L=10^8$ h$^{-2}$ L$_{\sun}$ to $10^{11}$ h$^{-2}$ L$_{\sun}$. These probabilities show 
a peak at low masses, associated with central galaxies, and a tail towards higher masses, associated with
satellites of the same luminosity. As the mass scale is increased, the peak related to central galaxies
broaden since the $L_c(M)$ relation increases slowly with increase in mass such that
one encounters a fractionally higher mass range with an increase in
luminosity.  Fainter late-type galaxies are in low mass halos, while brighter early type galaxies
are in galaxy groups and clusters.

By integrating the conditional probability distribution functions over luminosities,  we can calculate the
the probability distribution  of mass associated with the 2dF LF (van den Bosch  et al. 2003):
\begin{equation}
P(M) \; dM = \frac{\int \Phi(L|M) dL}{\int \Phi(L) dL} \frac{dn}{dM} dM \, .
\end{equation}
As shown in Figure~7, the LF is dominated by galaxies that occupy dark matter halos in the mass ranges
between $10^{10}$ M$_{\sun}$ to $10^{11}$ $M_{\sun}$.  While central galaxies dominate statistics in this mass range,
satellite galaxies become the dominant contributor to the LF from each mass scale. This, however, does not
imply that  at each luminosity, satellites dominate, but rather, as Figure~4 shows, central galaxies dominate.
Note that the behavior of central galaxies is similar to the average, but
on the other hand, early type or red galaxies, are primarily in halos with masses above $10^{11}$ M$_{\sun}$.

\subsection{Galaxy Bias}

Another useful quantity to compare with observed data is the galaxy bias, as a function of the luminosity. Using the conditional
LFs, we can calculate these as
\begin{equation}
b(L) = \int b_{\rm halo}(M) \frac{\Phi_i(L|M)}{\Phi_i(L)} \frac{dn}{dM} dM \, ,
\end{equation}
where $b_{\rm halo}(M)$ is the halo bias with respect to the linear density field (Sheth, Mo \& Tormen 2001; also,
Mo et al. 1997) and $i$ denotes the galaxy type.
In Figure~11, we show the galaxy bias as a function of the luminosity. We also divide the sample to galaxy
types. 

While the bias factors have similar shapes,
early-type galaxies are biased higher relative to the late-type galaxies; the difference in
the bias factor between early- and late-type galaxies is at the level of $\sim 5$\%. 
At low luminosities, the total sample bias is close to that of late-type galaxies, while at the bright-end
average bias  for the whole sample is close to that of early-type galaxies.
We also note an important differences in the bias when comparing satellite galaxies and central galaxies;
Satellite galaxies, on avearge, have a higher bias factor at all luminosities when compared to central galaxies.
This is due to the fact that satellite galaxies are preferentially in higher mass halos that  are, on averge,
biased higher with respect to the linear density field.
The average bias factor, for the whole
sample, however is dominated by central galaxies, for same reasons the LF is also dominated by central galaxies.
Bias measurements as a function of galaxy type exist in the form
of clustering information such as the correlation  length as a function of luminosity and type
(Norberg et al. 2002b). While we have not attempted to convert this clustering information to
obtain bias $b(L)$ as a function of galaxy type here, since it involves an additional
step of modeling, our next improvement in this approach is to compute clustering statistics, 
in which case a direct comparison could easily be made.  This clustering information has been
used in van den Bosch et al. (2003) when modeling the CLFs appropriate for the 2dFGRS survey.

\begin{figure}
\centerline{\psfig{file=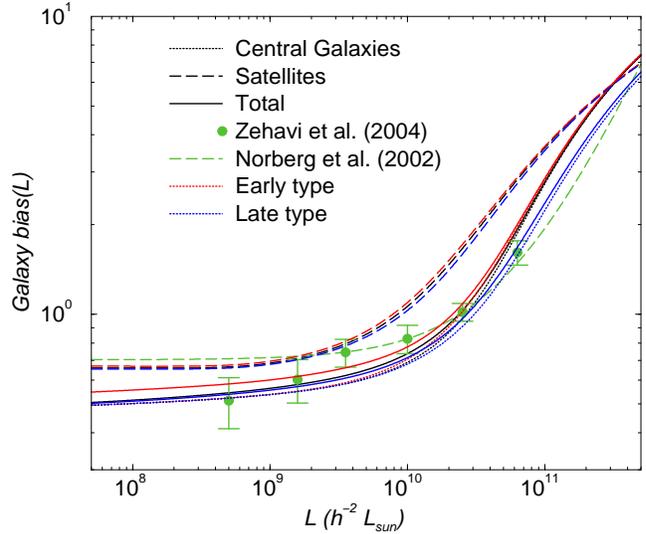,width=\hssize,angle=-90}}
\caption{Galaxy bias as a function of luminosity calculated from conditional
luminosity functions ({\it solid line}); 
we include the separate contributions from central galaxies ({\it dotted line}) 
and satellites ({\it dashed line}). Also shown are the SDSS of Zehavi et al. (2004), 
and a fit to the galaxy bias in 2dF data of Norberg et al. (2002b) ({\it dot-dashed line}). 
We also show the bias for galaxy types (early and late). Late-type galaxies are
expected to be in low-dense regions and their bias factor, relative to early type galaxies,
would be lower. Satellite galaxies, regardless of the type, are in more massive halos
and, thus, have higher bias factors relative to central galaxies. The average bias factor, for the whole
sample, however is dominated by central galaxies, for same reasons the LF is also dominated by central galaxies.}
\end{figure}

\section{Environmental Luminosity Functions}

Having discussed average statistics, we now focus on the LFs measured by Croton et al. (2004) as a function of the 
galaxy overdensity, $\delta_{\rm gal}$. These overdensities correspond to a volume of radius
8 $h^{-1}$ Mpc, and are measured following $\delta_{\rm gal} = (N_g -\bar{N}_g)/\bar{N}_g$
such that $-1 \geq \delta_{\rm gal}$.

To calculate the environmental LFs given by $\Phi(L|\delta_{\rm gal})$, the LF given $\delta_{\rm gal}$, following
Mo et al. (2004), we make one  important assumption;
We assume that the CLFs, $\Phi(L|M)$ are independent of the environment. The motivation for such an
assumption is inherent in the halo model (Cooray \& Sheth 2002), where one makes the assumption that galaxy distribution can
be associated with the halo mass rather than the environment. Similarly, observational studies based on
the Sloan survey
indicate that galaxy color may also be independent of the environment, implying that the CLFs for
galaxy types are not dependent on $\delta_{\rm gal}$ (Blanton et al. 2004). Thus,
given that $\Phi(L|M)$ is $\delta_{\rm gal}$ independent, all variations in $\Phi(L|\delta_{\rm gal})$
must come from variations in the halo mass function, as a function of $\delta_{\rm gal}$.
The halo mass function, in fact, is $\delta_{\rm dm}$ dependent; the peak background split (Sheth \& Tormen 1999,2002)
makes use of this dependence to extract information on, for example, relative biasing of the halos,
as a function of mass, to calculate halo bias factors. 
 
Thus, to calculate the LF as a function of the environment, as defined by the galaxy overdensity,
we define
\begin{equation}
\Phi_i(L|\delta_{\rm gal}) = \int_0^\infty \Phi_i(L|M) \frac{dn(M|\delta_{\rm gal})}{dM} dM \,,
\end{equation}
where $dn(M|\delta_{\rm gal})/dM$ is the conditional halo mass function.
Unfortunately, analytical techniques are not
adequate enough to reliably calculate the conditional mass function, $dn/dm(M|\delta_{\rm gal})$ given the
galaxy overdensity, $\delta_{\rm gal}$ (see, e.g., Mo et al. 2004 and discussion below).
Here, making  use of the assumption that CLFs are independent of $\delta_{\rm gal}$, we use
observed measurements of $\Phi(L|\delta_{\rm gal})$ to  extract information on $dn(M|\delta_{\rm gal})/dM$,
the conditional mass function of dark matter halos as a function of the galaxy overdensity.
In this approach, $dn(M|\delta_{\rm gal})/dM$ necessary to reproduce Croton et al. (2004) data can potentially be compared with
either improved models of the conditional mass function. While this comparison is beyond the scope of this paper, 
as it involves understanding certain aspects of the mass function, we plan to return to this topic later.
Later in this Section, we will, however, comment on the void mass function or the conditional mass function
when $\delta_{\rm gal} \sim -1$. Since Croton et al. (2004) measurements include the LF of galaxies in voids,
we can directly establish the void mass function using our modeling technique.

To extract information on $dn(M|\delta_{\rm gal})/dM$, motivated 
by numerical simulation based measurements in Mo et al. (2004) of the
conditional mass function, we assume one can model this as
\begin{equation}
dn(M|\delta_{\rm gal})/dM = A(\delta)\left[1+\frac{M}{10^{13}}\right]^{-\alpha(\delta)}dn/dM 
\label{eqn:dndmdelta}
\end{equation}
where $A(\delta)$ and $\alpha(\delta)$ are two parameters we will extract from the data given the
average mass function $dn/dM$ following the ST mass function.

These models fits to the Croton et al. (2004) data are shown in Figure~8.  For comparison, we also show the
average luminosity function (shown in Figure~4). Our model fits generally describe the data,
except in overdense regions, such as when $\delta_{\rm gal} > 6$, when we underpredict the
abundance of early type galaxies and overpredict the abundance of late-type galaxies.

The conditional mass functions that describe these environmental
luminosity functions are shown in Figure~9, where we plot separately the mass functions and the ratio of
mass functions to the average ST mass function. Under dense regions show a sharp decrease in the abundance
of high mass halos, while the over dense regions are such that it is a scaled version of the
ST mass function, but with an increasing abundance of high mass halos. 
In principle, these mass functions should be described by the conditional mass functions
that are used to calculate merger trees of dark matter halos or based on the peak-background
split (Sheth \& Tormen 1999, 2002). However, the require modification to the
mass function with $\delta \rightarrow \delta_c - \delta_{L,m}$ and $\sigma^2(M) \rightarrow \sigma^2(M)-\sigma^2(M')$ 
did not produce shapes of the extracted mass functions. In this replacement,
$\delta_c \sim 1.686$ is the standard overdensity for collapse, 
$M' = 4/3\pi R^3 \bar{\rho}_m (1+\delta_{m})$, and $\delta_{m}$ is the overdensity in mass
that corresponds to the overdensity in galaxies, and $\delta_{L,m}$ is the linearized overdensity corresponding to the
mass overdensity. The main reason for the difficulty in
obtaining $dn(\delta_{\rm gal})/dm$ is that we do not yet have
an accurate model for the relation between $\delta_{\rm gal}$, which is non-linear,
and $\delta_{m}$ and  $\delta_{L,m}$. In fact, these relations capture the biasing
of galaxies with respect to the linear density field. In Mo et al. (2004), authors
used simulations to estimate the environmental mass functions, as a function of the galaxy overdensity,
which were then compared with same LFs  directly measured in mock catalogs.
Here, we use observed data to extract information on the conditional mass functions.
We find reasonable agreement with the mass functions plotted in their Figure~2 \footnote{This is true
only if the labels in Figure~2 of Mo et al. (2004) is reversed from what is labeled there.
As shown in our figure~10, under dense mass functions show a low abundance of halos rather than the increase
abundance, relative to the over dense regions,  as suggested in Figure~2 of Mo et al. (2004). This is
likely to be a misprint.}

To further understand the dependence of these LFs on halo masses, we calculate conditional
probability for halos of mass $M$ to host galaxies given the environmental overdensity, $\delta_{\rm gal}$.
These probabilities are calculated from
\begin{equation}
P(M|\delta_{\rm gal}) = \frac{\int \Phi(L|M) dL}{\int \Phi(L) dL} \frac{dn(M|\delta_{\rm gal})}{dM} \, .
\end{equation}
In Figure~10, we plot $P(\log M|\delta_{\rm gal})$. These probabilities  show  mass scales  that
are important for the environmental LF, as a function of the galaxy overdensity.  In addition 
to the total distribution, we also show probabilities in terms of the galaxy type, and separated to both central and
satellite galaxies. The halos that contribute to the
LF in under dense regions host a higher fraction of late-type galaxies relative to early type ones. On the other hand,
in over dense regions, one finds both late-type and early-type galaxies,
with a large fraction of early-type galaxies coming from more massive halos  or galaxy 
groups and clusters, while the late-type fractional contribution is dominated by
galaxies in low mass halos.

To compare with measurements in Croton et al. (2004), we calculate one more quantity involving the mean
luminosity per galaxy, as a function of the galaxy density environment:
\begin{equation}
\frac{\bar{\rho}_L}{\bar{N}_g} = \frac{\int_{L_{\rm min}}  \Phi(L|\delta_{\rm gal}) L dL}{\int_{L_{\rm min}}  \Phi(L|\delta_{\rm gal}) dL} \, .
\end{equation}
Here, $L_{\rm min}$ was set at an absolute magnitude of $-17$ following the  measurements of Croton et al. (2004).
The Croton et al. (2004) data  and the same quantity based on model fits in Figure~8 are plotted in Figure~12.
They show general agreement, except in high dense environments, the mean luminosity per galaxy, in the case of
early type galaxies, is some what higher in our
models when compared to the data. This is due to the fact that our models under predict the abundance of
early type galaxies in dense environments, making them brighter on average than observed.

\begin{figure}
\centerline{\psfig{file=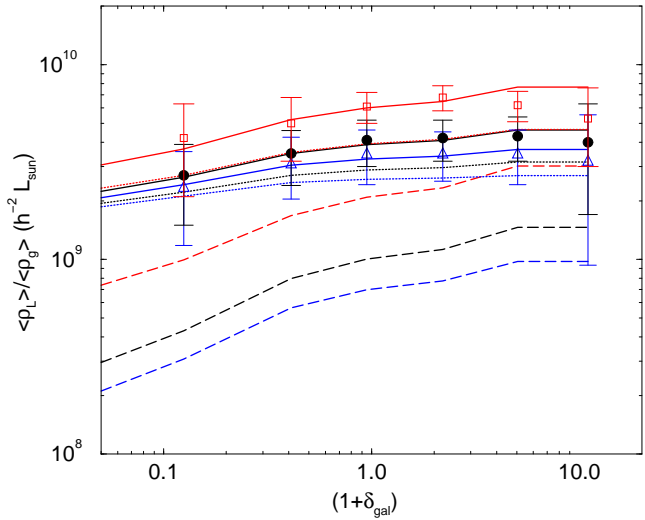,width=\hssize,angle=-90}}
\caption{The mean luminosity per galaxy as a function  of the environment, as measured by $\delta_{\rm gal}$. The data plotted are
direct measurements established by Croton et al. (2004) based on the measured environmental LFs, while the lines show the predictions based on
our model fits shown to describe the same environmental LFs (Figure~8). The solid circles show the total galaxy sample, 
red squares show early type galaxies, and blue triangles show late-type galaxies. We find a general agreement, except at the high dense regions, where
the mean luminosity per galaxy for early type galaxies is bit higher in our models when compared to the data. This is due to the fact that in our models,
we underpredict the abundance of early type galaxies, relative to observed LF.}
\end{figure}

\subsection{Galaxies in Voids}

As a final use of data in Croton et al. (2004), we use their environmental LF in 
{\it extreme voids}, with $-1 \leq \delta_{\rm gal} \leq -0.9$, to extract information on the
conditional mass function of the same environment. Our results our summarized in Figure~13. 
Note that the void LF show a sharp decrease in the abundance of galaxies with
absolute magnitudes brighter than -19. Since our conditional LFs are
independent of $\delta_{\rm gal}$, this puts a strong constraint on the void mass
function at the high mass end. Since luminosities grow with 
mass, following the $L_{\rm c}(M)$ relation, to restrict galaxy luminosities to be fainter than
the observed cutoff requires that no halos with masses greater than 10$^{13}$ M$_{\rm sun}$
be present in voids with a fractional abundance $M dn/dM$ greater than $10^{-8}$ h$^{3}$ Mpc$^{-3}$.

Given that the void mass function has received some attention in the literature,
we can make a direct comparison with our estimate of the mass function with previous
estimates to the extent analytical formulae are available. 
In Figure~13, we show several comparison drawn from published attempts to analytically
model the void mass function (also, see Gottlober et al. 2003). 
The methods by Goldberg et al. (2004) and Patiri et al. (2005)
generally fail to describe the void mass function both at the low- and high-end of
the mass function, though the abundance at a halo mass of $10^{11}$ M$_{\sun}$
is generally produced. The Sheth \& van de Weygaert (2003) analytic description for the
void mass function comes closest to describing the mass function required by the 2dF extreme
void LF. In fact, Sheth \& van de Weygaert (2003) description
managed to capture the low-end turn over in the void mass function, which is simply
associated with our efficiency function $f_c(M)$. This model, however, overproduces the
abundance of halos at the high mass end with a prediction for the presence of
$\sim$ $10^{14}$ M$_{\sun}$ halos with the same abundance as halos of $\sim$ $10^{13}$ M$_{\sun}$
required by the 2dF LF. The dashed lines in Figure~13(a) show the predicted LF based on the
Sheth \& van de Weygaert (2003) void mass function. The bright-end is over populated and the turnoff
in the LF moves to a higher luminosity than seen in the data.
Our suggestion for the void mass function, based on the Croton et al. (2004) {\it extreme void} LF,
may provide a useful guidance to analytically calculate the void mass function.
As part of an attempt to understand the conditional mass functions, we hope to return to this issue again
in the future.

\begin{figure*}
\centerline{\psfig{file=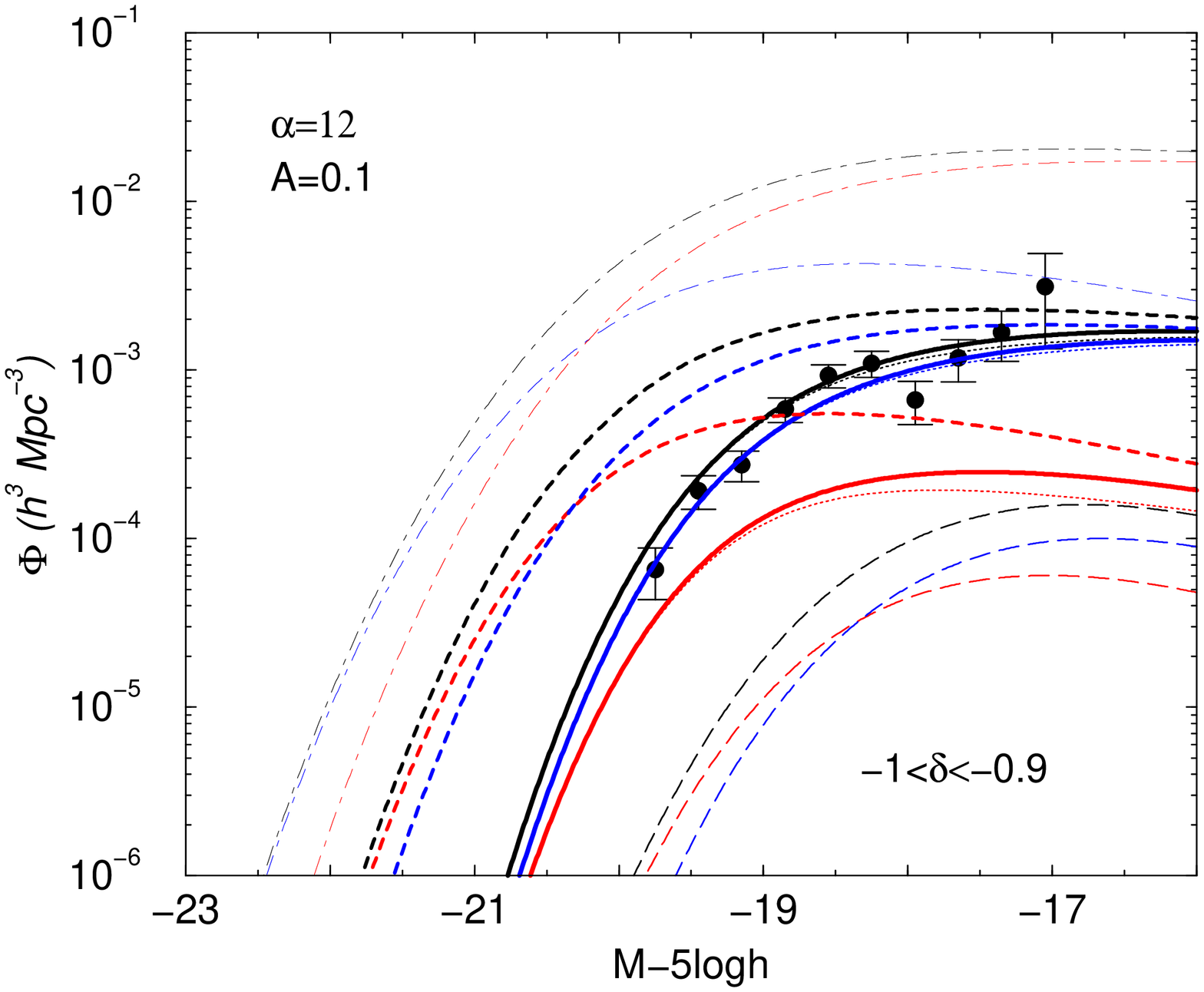,width=\hssize,angle=-90}
\psfig{file=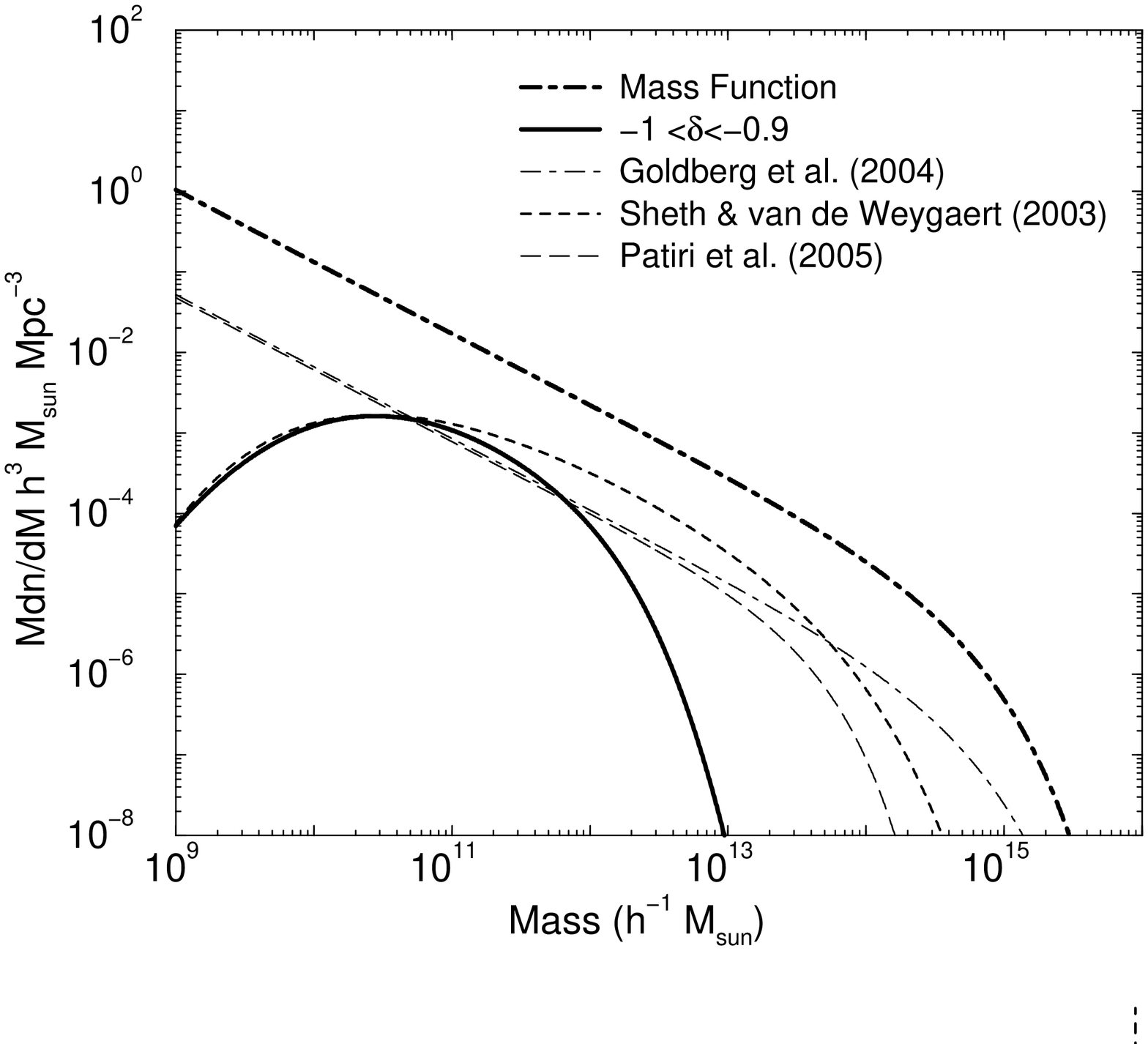,width=\hssize,angle=-90}}
\caption{A comparison of the void LF (a) and the void mass function (b). Here, we show the
galaxy LF when $-1 <\delta_{\rm gal} < -0.9$, where $\delta_{\rm gal}$ is measured over
a radius of 8 $h^{-1}$ Mpc. The data points are measurements from
Croton et al. (2004), where only the total LF in voids was measured. 
The curves follow the same line-styles as Figure~8, and with dot-dashed lines
showing the mean LF in Figure~4. In (b), we show the mass function that fits the Croton et al.
(2004) data with a thick solid line. The dot-dashed line is the ST mass function.
We also show three predictions from the literature on the void mass function.
Both Goldberg et al. (2004) and Patiri et al. (2005) fail to describe both the low- and high-mass
end shapes of the mass function derived here. Interestingly, the analytic
description by Sheth \& van de Weygaert (2004) fits the low-mass end of void mass function,
which is captured by our function $f_c(M)$ applied to the average mass function,
but over predicts the abundance of high mass halos in voids. For example, the void LF
calculated with the Sheth \& van de Weygaert (2004) mass function for voids is shown
in (a) with dashed-lines. While the low-luminosity end agrees with the data, the high luminosity end,
and the turnover in the luminosity function is higher than suggested by the data. This increase
is associated with the high mass allows in the  Sheth \& van de Weygaert (2004) void mass function.}
\end{figure*} 

\section{Discussion and Summary}

In this paper, we have made use  of the scaling relation between the central and the total galaxy luminosity 
of a dark matter halo as a function of the halo mass,  to construct the CLF of galaxies.
The CLF provides a powerful technique to understand galaxy properties, especially 
on the spatial distribution  of galaxies, in the era of wide-field large scale structure surveys 
where large galaxy samples can easily be subdivided with adequate statistics (Yang et al. 2003b).
The CLF is closely related to the halo occupation number, which is the average
number of galaxies given the halo mass (Cooray \& Sheth 2002). 
The halo occupation number captures galaxy
statistics that simply treat each galaxy in the sample equally and is useful when attempting
to understand statistics such as galaxy clustering statistics, on average.
Higher order statistics of the galaxy distribution require detailed statistics of halo occupation beyond the mean.
On the other hand, galaxies vary in luminosity and color. Therefore, a  more useful quantity to consider
is the conditional occupation number, ie., the number of galaxies in a given halo mass with given luminosity.
These conditional occupation numbers are in fact the conditional luminosity functions we have
described here.  

When compared to previous descriptions of the CLF in the literature (e.g., Yang et al. 2003b; Yang et al. 2005),
we make several improvements by dividing the galaxy sample to galaxies that  are in centers of dark matter halos 
(central galaxies),
and in subhalos of a main halo (satellite galaxies). This division is central to are arguments on how the CLF
is shaped and assumes that different physics govern the evolution of these two components. In our underlying physical
description, following Cooray \& Milosavljevi\'c (2005a), central galaxies grow in luminosity through dissipationless
merging of satellite galaxies. These satellites are both early- and late-types, with the
fraction of early type galaxies taken to be dependent on both the halo mass and the satellite galaxy luminosity. 
Underlying physical reasons for this dependence is not clear, but could very well associated with a tidal stripping effect
that removes gas and the stellar content of late-type galaxies in more massive halos and convert these galaxies to early-types.
It will be helpful to understand if numerical and semi-analytic models predict the fractions of late to early type galaxies,
including the suggestion that the fraction in satellites changes at a luminosity around $\sim 5 \times 10^9$ L$_{\sun}$
from more late-types at lower luminosities than this value to more early-types.

Our simple empirical model describes the conditional luminosity functions of galaxy types,
as well as the total sample, measured with the 2dF galaxy group catalog from cluster to group mass scales 
by Yang et al. (2005). A comparison of our CLFs to ones in Yang et al. (2005; their Figure~9 and 11)
reveal that our model can account for the peak  of the CLF at high luminosities; this peak is associated with 
central galaxies, which we have modeled with a log-normal distribution. Similar peaks are also observed in
the conditional baryonic mass function of stars in semi-analytical models of galaxy formation (Zheng et al. 2004) and
in the luminosity functions of clusters where all galaxies are included (Trentham \& Tully 2002).
The slope of the satellite CLF changes from -1 at cluster mass scales to 0 at group mass scales.
Again this could be a reflection of the merging evolution of satellite galaxies as hierarchical merging builds up bigger halos.
It will be interesting to see if a simple extension of the analytic model of Cooray \& Milosavljevi\'c (2005a), which involves dynamical friction, may explain the change in the slope as parent halo mass is increased.

Given the observational measurements of the LF as a function of the
environment using 2dF data by Croton et al. (2004), as measured by the galaxy overdensity measured over volumes
corresponding to a radius of 8 h$^{-1}$ Mpc, we extended our CLF model to
describe environmental luminosity functions. While it would have been useful to have direct predictions that
can compare with observations, we failed to do this due to lack of information on the conditional mass function,
or the mass function  of dark matter halos given the galaxy overdensity. In Mo et al. (2004) predictions were made
using the conditional mass function measured in numerical simulations. Here, we use Croton et al. (2004)
measurements to establish information on the conditional mass functions, from environments such as galaxy voids to
dense regions. With these mass functions, we also estimate statistical quantities as
probability distribution function of halo mass, as a function of the galaxy overdensity. We find that
the preferred environment of blue galaxies are underdense regions in low mass halos, while the
early-type, red galaxies are mostly in dense environments and dominated by satellites of larger mass halos.
The shapes of the mass functions we have extracted could eventually be compared to numerical simulations or
analytical techniques.

Using Croton et al. (2004) LF for galaxies in voids, we also establish the void mass function.
We find that the void mass function is peaked at halo masses around $\sim 10^{11}$ M$_{\sun}$.
Such a peak in the void mass function is predicted in the analytical calculation by Sheth \& van de Weygaert (2003).
We do not, however, find a tail to higher halo masses as suggested by the description in
 Sheth \& van de Weygaert (2003). The void mass function must sharply turn over; if not, one
would predict brighter galaxies in void environments than suggested by the measurements in
Croton et al. (2004).

To summarize our paper, main results are:

(1)  The galaxy LF, which is an average of CLFs over the halo mass function, is primarily shaped by the
$L_{\rm c}(M)$ relation; the faint-end slope of the LF reflects the faint-end scaling of the $L_{\rm c}(M)$
relation while the bright-end turn off in the LF, generally described by the exponential
cut off in the Schechter (1976) fitting form, is determined by the scatter in the $L_{\rm c}(M)$ relation.
Understanding the $L_{\rm c}(M)$ relation and it's scatter is central to understanding the galaxy LF.

(2) The fraction of galaxy types as a function of the halo mass (Figure~1). The galaxy types are
distributed such that one finds essentially all late-types in low mass halos and mostly
early-types in high mass halos. In the case of early-types, the fraction is dominated by satellite galaxies
rather than central galaxies in halo centers. These fractions may capture interesting physics such
as tidal stripping that happens in dense environments and could also be responsible for the fractional
change of galaxy types, from dominate late-type to early-type, as the luminosity of satellite galaxies
is increased.

(3) The mass dependent slope for the satellite CLF, where $\gamma(M) \sim -1$ at galaxy cluster mass scales and
$\gamma \sim 0$ at poor galaxy group mass scales. The slope may be a reflection of the hierarchical merging process
and predictions on $\gamma(M)$ do not yet exist in the literature.

(4) The conditional mass functions (Figure~9), or the mass function of dark matter halos given the galaxy overdensity
measured over a volume corresponding to a  radius of 8 h$^{-1}$ Mpc. We will leave it as a challenge to improve
analytical techniques produce the required mass functions to compare with what we suggest is needed to
explain Croton et al. (2004) measurements. Any disagreements, if understood, may suggest that
our central assumption that the CLF is independent of the galaxy overdensity is incorrect.
If that's the case, galaxy formation and evolution involves additional parameters beyond the halo mass
and could question the viability of the halo approach to describe galaxy statistics.

(5) Galaxy bias predictions as a function of luminosity given the galaxy type (Figure~11).
While our predictions agree with the bias predicted for total samples, it will be useful to
understand clustering bias as a function of galaxy color as well. In an upcoming paper,
we will extend the CLFs developed here to describe clustering properties of galaxies
and a direct comparison to measurements in Norberg et al. (2002b).

{\it Acknowledgments:} 

We thank Milos Milosavljevi\'c for his contributions that began this project from an attempt to
understand a simple observed relation (Cooray \& Milosavljevi\'c 2005a) and
Frank van den Bosch for his detailed comments and suggestions that convinced the author to use this
simple relation to model the LF. The author also thanks Darren Croton and Xiaohu Yang for providing
electronic files of various measurements related from 2dFGRS that are modeled in the paper.
The author thanks members of Cosmology and Theoretical Astrophysics groups
at Caltech and UC Irvine for useful discussions.

\end{document}